\documentclass[                           %
reprint,aps,pre,showpacs,                 %
superscriptaddress,nofootinbib            %
]{revtex4-1}                              %
\usepackage{amsthm,amssymb,amsmath}       %
\usepackage{graphicx}                     %
\usepackage{longtable,comment}            %
\usepackage[colorlinks = true,            %
            linkcolor = blue,             %
            urlcolor  = blue,             %
            citecolor = blue,             %
            anchorcolor = blue]           %
            {hyperref}                    %

\begin{document}

\title{Deformed Fokker-Planck equation:
       inhomogeneous medium with a position-dependent mass}

\author{Bruno G.\ da Costa}
\email{bruno.costa@ifsertao-pe.edu.br}
\affiliation{Instituto Federal de Educa\c{c}\~ao, Ci\^encia e Tecnologia do
             Sert\~ao Pernambucano,
             Rua Maria Luiza de Ara\'ujo Gomes Cabral s/n,
			 56316-686 Petrolina, Pernambuco, Brazil}
\author{Ignacio S.\ Gomez}
\email{nachosky@fisica.unlp.edu.ar}
\affiliation{Instituto de Fisica, Universidade Federal da Bahia,
             R.\ Barao de Jeremoabo s/n, 40170-115 Salvador, Bahia, Brazil}
\affiliation{National Institute of Science and Technology for Complex Systems,
             Brazil}
\author{Ernesto P.\ Borges}
\email{ernesto@ufba.br}
\affiliation{Instituto de Fisica, Universidade Federal da Bahia,
             R.\ Barao de Jeremoabo s/n, 40170-115 Salvador, Bahia, Brazil}
\affiliation{National Institute of Science and Technology for Complex Systems,
             Brazil}
\date{\today}
\begin{abstract}
We present the Fokker-Planck equation (FPE) for an inhomogeneous medium
with a position-dependent mass particle by making use of the Langevin equation,
in the context of a generalized deformed derivative
for an arbitrary deformation space where the linear (nonlinear) character
of the FPE is associated with the employed
deformed linear (nonlinear) derivative.
The FPE for an inhomogeneous medium with a position-dependent
diffusion coefficient is equivalent to a deformed FPE
within a deformed space, described by generalized derivatives,
and constant diffusion coefficient.
The deformed FPE is consistent with the diffusion
equation for inhomogeneous media when the temperature
and the mobility have the same position-dependent
functional form as well as with the nonlinear Langevin approach.
The deformed version of the $H$-theorem permits to express 
the Boltzmann-Gibbs entropic functional
as a sum of two contributions, one from the particles and the other
from the inhomogeneous medium.
The formalism is illustrated with the infinite square well and the
confining potential with linear drift coefficient.
Connections between superstatistics and
position-dependent Langevin equations are also discussed.
\end{abstract}
\maketitle

\section{\label{sec:intro}Introduction}

Diffusion is understood as the thermal motion
of particles, which is macroscopically translated into a net flux
from one region to another.
The standard way to quantify this phenomenon is to consider
that the particles are subjected to drag (properly of the fluid)
and random (Brownian motion) forces,
which gives place to the Langevin equation \cite{Langevin-1908}.
To link this classical description with a probabilistic characterization,
the usual strategy is to rewrite the Langevin equation
in terms of the probability density function (PDF),
thus obtaining the Fokker-Planck equation (FPE) \cite{Risken}.
The FPE has been widely investigated in the literature, mainly
applied to the study of different types of diffusion,
including the normal and anomalous ones (associated to linear and
nonlinear FPE)
\cite{Scher-Montroll-1975,Plastino-Plastino-1995,Tsallis-Bukmann-1996,
Borland-PRE-1998,Borland-PLA-1998,DeVoe-2009}.
Subsequent applications in multiple kinds of phenomena have displayed
the relevance of the FPE in the field of statistical physics
\cite{Frank-2005,Muskat,
Klemm-Muller-Kimmich-1997,
Barrozo-Moreira-Aguiar-AndradeJr-2009,
Tsallis-1988,
Tsallis-2009,
Andrade-Silva-Moreira-Nobre-Curado-2010}.
In particular, the FPE in a specific medium
have presented an increasing interest since it allows to characterize
electron diffusion \cite{Hizanidis-1989},
photoinduction in nonequilibrium processes \cite{Jang-2016},
rarefied gases and heterogeneous media
\cite{Suciu-Radu-Attinger-Schuler-Knabner-2015,
     Collyer-Connaughton-Lockerbyco-2016},
interfaces-membranes \cite{Grassi-Raudino-2014},
multiple diffusion from fractional kernel operators \cite{Maike-2018},
superfast diffusion in porous media \cite{Xu-2019},
among others.

In addition, theoretical investigations
have shown an intimate connection between generalized FPE,
$H$-theorem,
master equations and entropic forms,
highlighting the role played by the nonextensive statistics
\cite{Frank-2001,
      Schwammle-Curado-Nobre-2007,
      Ribeiro-Nobre-Curado-2011,
      Sicuro-Rapcan-Tsallis-2016}.
Along to this progresses, the mathematical structure inherited
by nonextensive statistics turned out to be a useful tool
to generalize concepts of statistical mechanics.
Some mathematical structures have been presented
\cite{Nivanen-Mehaute-Wang-2003,
      Borges-2004,
      Tsallis-1994,
      Lobao-Cardoso-Pinho-Borges-2009,
      Tempesta-2011},
referred to as generalized algebras.

Parallel to this developments,
the research on systems with a position-dependent effective mass emerged
for describing transport phenomena in semiconductors heterostructures
\cite{Bastard-1975,vonRoos-1983}
provided with a position-dependent chemical composition.
The starting point of this approach was the Wannier-Slater theorem
for the wave function of the conduction band in
homogeneous semiconductors, from which its extension to an inhomogeneous one
led to several ways for defining the kinetic energy operator
\cite{vonRoos-1983}.
This ambiguity, called the ordering problem,
was unified together with the requirement of hermiticity
by von Roos \cite{vonRoos-1983}.
Recently, from a particular case of the von Roos kinetic energy operator,
a deformed Schr\"odinger equation for position-dependent mass has been studied
\cite{CostaFilho-Almeida-Farias-AndradeJr-2011,
      CostaFilho-Alencar-Skagerstam-AndradeJr-2013,
      Barbagiovanni-2014,
      Costa-Borges-2014,
      Costa-Borges-2018,
      Costa-Gomez-2018,
      Nascimento-Ferreira-Aguiar-Guedes-CostaFilho-2018,
      Costa-Gomez-Santos-2020}
and linked with a generalized translation operator inherited
by the generalized $q$-algebra
\cite{Nivanen-Mehaute-Wang-2003,
      Borges-2004,
      Tsallis-1994,
      Lobao-Cardoso-Pinho-Borges-2009,
      Tempesta-2011}.
Position-dependent mass systems have been proven to be a useful
theoretical tool in multiple areas and fairly fitting to experimental data:
density functional theory \cite{Bencheikh-2004},
supersymmetric quantum mechanics \cite{Ioffe-2016},
nuclear physics \cite{Alimohammadi-2017},
nonlinear optics \cite{Li-2017},
Landau quantization \cite{Mustafa-2020},
among others.

The goal of this paper is to present the FPE
for an inhomogeneous medium with a variable diffusion coefficient
within the position-dependent mass scenario
\cite{CostaFilho-Almeida-Farias-AndradeJr-2011,
      CostaFilho-Alencar-Skagerstam-AndradeJr-2013,
      Costa-Borges-2014,
      Costa-Borges-2018,
      Costa-Gomez-2018},
by means of a generalized deformed derivative,
where the deformation of the space
univocally determines the mass as well as
the dumping and the diffusion coefficients.
As a consequence, we find an equivalence between the FPE
in an inhomogeneous medium and a deformed FPE
with constant mass and constant diffusion coefficient.
In particular, we analyze the deformed FPE that results from the
$q$-algebra
\cite{Nivanen-Mehaute-Wang-2003,Borges-2004},
controlled by a real and continuous dimensionless parameter $q$.
The solutions exhibit an asymmetric spatial distribution that
physically corresponds to the inhomogeneity of the medium.
We present a generalized version of the $H$-theorem
in which the total entropy is the sum of
the Boltzmann entropy with
an additional term associated
to the inhomogeneity of the medium.
The deformed FPE results compatible with the van Kampen's approach
for inhomogeneous diffusion \cite{vanKampen-87,vanKampen-book,Landauer},
when the temperature and the mobility have the same position-dependent
functional form, with the superstatistics version of the Langevin equation
\cite{vanderStraeten,superstatistics}
and also with the nonlinear Langevin equation \cite{vanKampen-book}.

The work is structured as follows.
In Section \ref{sec:linear-nonlinear-FPE}
we review
the FPE construction from Langevin equation
along with diffusion in inhomogeneous media \cite{vanKampen-87,vanKampen-book}
and the $q$-algebra.

Section \ref{sec:FP-inhomogeneous-media}
is devoted to generalize the FPE for an inhomogeneous medium
(the deformed FPE)
from its corresponding Langevin equation, by employing a
generalized derivative operator determined by the
position-dependent mass function
and the properties of the medium.
Given an arbitrary deformation space, we begin
by defining a deformed linear derivative and its
associated nonlinear dual derivative,
and then we establish a link between the linearity (nonlinearity)
of the equation expressed
by the deformed space and the
deformed linear
derivative (dual nonlinear
derivative) used.
We also present a generalized version of the $H$-theorem
for the FPE in a general deformed position space,
and the equivalence of the deformed FPE with the nonlinear Langevin approach
\cite{vanKampen-book}.

In Section \ref{sec:solutions} we specialize for the
case of the $q$-algebra inspired by nonextensive statistics,
and we obtain its associated $q$-deformed FPE,
as well as an analytical expression for the general solution
within the deformed space.

Next, in Section \ref{sec:applications}
we illustrate the formalism presented for two potentials:
the infinite square well
and the confining potential with linear drift coefficient.

Section \ref{sec:discussion}
is devoted to discuss the deformed FPE in some diffusive contexts:
the van Kampen's diffusion for inhomogeneous media
\cite{vanKampen-87,vanKampen-book},
the superstatistics of the Langevin equation
\cite{vanderStraeten,superstatistics}
and the anomalous diffusion in optical lattices \cite{Lutz,Beck-entropy}.
The van Kampen's diffusion equation can be expressed in terms of the deformed
FPE when the temperature and the mobility of the particle have the same
position-dependent functional form.
There is a connection between superstatistics
and position-dependent mass Langevin equations.
We indicate two possible fluctuation theorems
linked with position-dependent mass systems.
In the context of optical lattices, for the anomalous diffusion regime
we express the stationary Rayleigh equation of the Wigner distribution
as a deformed FPE.

Finally, in Section \ref{sec:conclusions}
some conclusions and perspectives are outlined.

\section{\label{sec:linear-nonlinear-FPE}
          Preliminaries}

We present a review of the Langevin and the Fokker-Planck equations,
the van Kampen's and superstatistics inhomogeneous diffusion
along with the $q$-calculus.

\subsection{\label{subsec:FPE-preliminaries}
            Langevin and Fokker-Planck equations}

A single particle of mass $m_0$ in a fluid of viscosity coefficient
$\lambda_0$ subjected to an external potential $V(x)$
(i.e., an external force $F(x) = -dV(x)/dx$)
and a random force $R(t)$ has an equation of motion
that can be obtained from the Lagrangian
\begin{equation}\label{eq:Lagrangian-Langevin}
 \mathcal{L}(x,\dot{x},t)=
 \frac{1}{2}m_0\dot{x}^2 -U(x,t)
\end{equation}
and using the Euler-Lagrange equation
\begin{equation}
 \label{eq:Euler-Lagrange}
 \frac{d}{dt}\left( \frac{\partial \mathcal{L}}{\partial\dot{x}} \right)
 - \frac{\partial \mathcal{L}}{\partial x}
 + \frac{\partial Q}{\partial \dot{x}} = 0,
\end{equation}
where $Q = \frac{1}{2} m_0 \lambda_0 \dot{x}^2$
is a Rayleigh dissipation function, and $U(x, t) = V(x) - xR(t)$
is the potential due to conservative and random forces.
Thus, the corresponding Langevin equation is
\begin{equation}
 \label{eq:Langevin-equation}
 \ddot{x}=-\lambda_0 \dot{x}+f(x)+\xi(t),
\end{equation}
with $f(x)=F(x)/m_0$ and $\xi(t)=R(t)/m_0$.

Generally, the Langevin equation for $N$ stochastic variables
$\vec{y} = \{y_1, ..., y_N\}$ with $M$ white Gaussian noises
$\vec{\xi} = \{\xi_1, ..., \xi_M\}$
and a diffusion coefficient $D_0$
 (i.e., $\langle \xi_j(t) \rangle = 0$ and
  $\langle \xi_j(t) \xi_l(t') \rangle = 2D_0 \delta_{jl}
  \delta (t' - t)$ $\forall j,l=1,\ldots,M$
  and $\forall t$)
is
\begin{equation}
\label{eq:generalized-langevin}
	\frac{dy_i}{dt} =
	A_i(\vec{y}, t) + \sum_{j} B_{ij} (\vec{y}, t) \xi_j(t),
	\quad (i=1,...,N),
\end{equation}
from which the diffusion equation results
\cite{Risken}
\begin{eqnarray}
 \label{eq:generalized-difusion}
 \frac{\partial P}{\partial t} &=&
  -\sum_{i} \frac{\partial}{\partial y_i}
    \left\{
     \left[
      A_i(\vec{y}, t) +
      \frac{\Gamma}{2} \sum_{jl} B_{jl}(\vec{y},t)
                        \frac{\partial B_{il} }{\partial y_j}
     \right] P
    \right\}
    \nonumber \\
    & & + D_0 \sum_{ij} \frac{\partial^2}{\partial y_i \partial y_j}
               \left\{
                \left[
                 \sum_l B_{il}(\vec{y},t) B_{jl}(\vec{y},t)
                \right] P
               \right\}.
\end{eqnarray}

In the overdamped limit of the Langevin equation
(i.e.~$\lambda_0\gg \tau^{-1}$ with $\tau$ a coarse-grained
time scale), the inertia term $\ddot{x}$
is negligible compared with
$\lambda_0 \dot{x}$, so
$dx/dt= [f(x)+{\xi}(t)]/\lambda_0$.
Substituting $y=x$, $A(x)=f(x)/\lambda_0$ and $B=1/\lambda_0$
in Eq.~(\ref{eq:generalized-difusion}) we obtain the
unidimensional FPE
\begin{eqnarray}
 \label{eq:fokker-planck}
 \frac{\partial P}{\partial t}
 = - \frac{\partial }{\partial x} \left[ A(x) P(x,t) \right]	
   + \frac{\Gamma}{2} \frac{\partial^2 P}{\partial x^2},
\end{eqnarray}
with $A(x)$ the confining potential and
$\Gamma/2 = D_0/\lambda_0^2$
a parameter related to the diffusion mechanism.
The general solution of Eq.~(\ref{eq:fokker-planck}) depends on the
confining potential and the initial conditions.
For long times ($t\rightarrow \infty$),
the solution of the FPE tends to the
stationary distribution
\begin{equation}\label{eq:stationary-sem-deformacao}
P^{(\textrm{st})}(x) = C\exp \left[\frac{2}{\Gamma}\int^x A(x')dx'\right],
\end{equation}
where $C$ is the normalization constant.
Analytical solutions are obtained for a few instances.
We briefly review two typical cases \cite{Risken}.
In absence of external forces $A(x)=0$ (free particle case)
with the initial condition $P(x,t=0)=\delta(x)$,
the probability distribution is a Gaussian
\begin{equation}
 P(x,t) = \frac{1}{\sqrt{2\pi \Gamma t}} e^{-x^2/(2\Gamma t)},
\end{equation}
corresponding to normal diffusion.
For a linear potential $A(x) = -\alpha x$ with the same initial condition
$P(x,t=0)=\delta(x)$ and the boundary conditions
$P(x,t)|_{x\rightarrow \pm\infty}
  = \partial P(x,t)/\partial x |_{x\rightarrow \pm\infty}
  = [A(x)P(x,t)]|_{x\rightarrow \pm\infty}=0 $
$\forall t$,
the solution is
\begin{equation}
 \label{generalsol-fokkerplanck}
  P(x,t) = \sqrt{\frac{\alpha}{\pi \Gamma (1-e^{-2\alpha t})}}
          \exp
              \left[
               {-\frac{\alpha x^2}{\Gamma (1-e^{-2\alpha t})}}
              \right],
\end{equation}
which tends asymptotically for $t\rightarrow \infty$
to the Gaussian stationary solution
\begin{equation}
\label{eq:estacionariasol-fokkerplanck}
 P^{(\textrm{st})}(x)={\sqrt{\frac{\alpha}{\pi \Gamma}}}e^{-\alpha x^2/\Gamma}.
\end{equation}

\subsection{\label{subsec:van-Kampen-superstatistics}
            Diffusion in inhomogeneous media:
            van Kampen's approach and Superstatistics
}

The diffusion equation for a single particle
immersed in an inhomogeneous medium with Brownian motion
and position-dependent mobility $\mu(x)$ and temperature $T(x)$,
whose phase space distribution obeys Kramers' equation
provided with a potential $V(x)$ that causes a drift velocity $\mu(x)V'(x)$,
is given by (denoted by $\rho(x,t)$ in \cite{vanKampen-87})
\begin{eqnarray}
\label{eq:van-Kampen-equation}
\frac{\partial P}{\partial t} &=&
		\frac{\partial}{\partial x}[\mu(x)V'(x)P(x,t)]+
		\nonumber \\
		&&
		\frac{\partial}{\partial x}
		\left\{ \mu(x)\frac{\partial}{\partial x}[T(x)P(x,t)]\right\}.
\end{eqnarray}
Its stationary solution is (equation (6) of \cite{vanKampen-87})
\begin{equation}\label{van-Kampen-stationary}
P^{(\textrm{st})}(x)
	= \frac{C}{T(x)} \exp \left[-\int^{x}
	  \frac{V'(x^{\prime})}{T(x^{\prime})}dx^{\prime}\right].
\end{equation}

Complementarily, superstatistics has proven to be a useful tool
for describing nonequilibrium steady-state of inhomogeneous systems
with spatial-temporal fluctuations of temperature
(or, more generally, fluctuations of any intensive quantity)
\cite{superstatistics}.
The system is conceived as composed of small
elementary cells in equilibrium in a small spatial-temporal scale
whose spatial correlation length is of the order of their sizes,
and the relaxation time is much smaller than their characteristic times,
thus their volumes are sufficiently large
for statistical mechanics to be locally valid
and
canonical ensemble applies.
In order to generalize the Langevin equation in the context of superstatistics,
in \cite{vanderStraeten} is assumed the set of equations
\begin{subequations}
 \begin{align}
  \label{eq:Langevin-superstatistics}
  &\frac{dv}{dt} =
   -\gamma v + \frac{F(x)}{m_0}
   +\sqrt{\frac{2\gamma}{m_0\beta(x)}} \xi(t),
  \\
  &\frac{dx}{dt} = v.
 \end{align}
\end{subequations}
$\beta(x)$ is the inverse of the temperature,
a position-dependent variable within this context,
with $\gamma$ the constant $\lambda_0$
and $\xi(t)$ now has a variance normalized.
In the overdamped limit of (\ref{eq:Langevin-superstatistics})
the associated Fokker-Planck equation for
the stationary distribution $P^{(\textrm{st})}$
results\footnote{Also studied by Borland \cite{Borland-PLA-1998}
          in connection with the Tsallis distribution
          by imposing specific conditions on $F(x)$ and $\beta(x)$.}
\begin{equation}
 \label{superstatistics-FPE}
0 = -\frac{\partial}{\partial x}[F(x)P^{(\textrm{st})}(x)]
    +\frac{\partial^2}{\partial x^2} \left[ \frac{P^{(\textrm{st})}(x)}{\beta(x)} \right]
\end{equation}
which constitutes a particular case of
van Kampen's equation (\ref{eq:van-Kampen-equation}) for
$\mu(x)=\mu_0=\textrm{constant}$ and $\beta(x)=\frac{1}{T(x)}$,
and whose stationary solutions are (equation (10) of
\cite{vanderStraeten})
\begin{equation}
\label{eq:superstatistics-stationary}
P^{(\textrm{st})}(x)=Z^{-1}\beta(x)\exp\left(\int^x F(x')\beta(x')dx' \right).
\end{equation}

In Section \ref{sec:discussion} we will return to
the van Kampen's approach and superstatistics' Langevin equation
(\ref{eq:Langevin-superstatistics})
in order to show the consistency and connections with the
position-dependent mass Langevin equation (\ref{eq:masslangevin-pdm}).

\subsection{\label{subsec:q-math}
            $q$-Deformed calculus }

Inspired by nonextensive statistics, the deformed
$q$-exponential and $q$-logarithm functions defined by
\cite{Tsallis-1994}
\begin{equation}
\exp_q (u) \equiv [1 + (1-q)u]_{+}^{1/(1-q)},
\end{equation}
and
\begin{equation}
\ln_q (u) \equiv \frac{u^{1-q}-1}{1-q} \quad (u > 0)
\end{equation}
have an associated nondistributive algebraic structure
\cite{Borges-2004,Nivanen-Mehaute-Wang-2003}
(called $q$-algebra):
the $q$-sum $a\oplus_{q} b = a+b+(1-q)ab$,
the $q$-difference $a\ominus_{q}b=\frac{a-b}{1+(1-q)b}$ ($b\neq \frac{1}{q-1}$),
the $q$-product $a\otimes_{q}b=[a^{1-q}+b^{1-q}-1]_{+}^{\frac{1}{1-q}}$
($a,b>0$),
and the $q$-ratio $a\oslash_{q}b = [a^{1-q}-b^{1-q}+1]_{+}^{\frac{1}{1-q}}$
($a,b>0$)
(with $[\cdot]_+ \equiv \max(\cdot,0)$).
From the $q$-difference a
deformed derivative $\mathcal{D}_q$ is defined as follows
\cite{Borges-2004}
\begin{equation}
 \label{eq:d_qu-diff}
 d_q u = \lim_{u' \to u} u' \ominus_q u=\frac{du}{1+(1-q)u},
\end{equation}
\begin{eqnarray}
 \label{eq:borges-linear-derivative-q-diff}
 \mathcal{D}_q f(u)
	&=& \frac{df(u)}{d_{q}u} \nonumber \\
	&=& \lim_{u' \to u} \frac{f(u')-f(u)}{u'\ominus_q u} \nonumber \\
	&=&[1+(1-q)u]\frac{df}{du},
\end{eqnarray}
along with its dual derivative,
\begin{eqnarray}
 \label{eq:borges-nonlinear-derivative-q-diff}
  \displaystyle \widetilde{\mathcal{D}}_{q} f(u)
        &=& \frac{d_{q}f(u)}{du} \nonumber \\
		&=& \lim_{u'\to u}\frac{f(u')\ominus_q f(u)}{u'- u} \nonumber \\
	 	&=& \frac{1}{[1+(1-q)f(u)]}\frac{df}{du},
\end{eqnarray}
where $d_q$ stands for a deformed differential.
In order to emphasize their features,
from now on the deformed derivative and its dual
will be indistinctly called as linear deformed derivative
and nonlinear deformed derivative, respectively.
The $q$-deformed calculus allows to recover the usual one for $q\rightarrow1$.
One of its applications concerns a generalization of the Schr\"odinger equation
for position-dependent mass systems
\cite{CostaFilho-Almeida-Farias-AndradeJr-2011,
      CostaFilho-Alencar-Skagerstam-AndradeJr-2013,
      Barbagiovanni-2014,
      Costa-Borges-2014,Costa-Borges-2018,
      Costa-Gomez-2018,
      Nascimento-Ferreira-Aguiar-Guedes-CostaFilho-2018,
      Costa-Gomez-Santos-2020}.
In fact, a $q$-deformed linear Schr\"odinger equation
is employed to describe systems with position-dependent mass
consistent with the von Roos kinetic energy operator \cite{vonRoos-1983}
\begin{eqnarray}
\label{eq:SE-pdm}
i\hbar \frac{\partial \Psi_q (x,t)}{\partial t}
		&=& -\frac{\hbar^2}{2m(x)} \frac{\partial^2 \Psi_q (x, t)}{\partial x^2}
	    \nonumber \\
		& &   -\frac{\hbar^2}{4} \frac{d}{dx} \left( \frac{1}{m(x)} \right)
			\frac{\partial \Psi_q (x, t)}{\partial x}	
		\nonumber \\
		& &	 + V(x) \Psi_q (x,t)
\end{eqnarray}
with
\begin{equation}
\label{eq:m(x)}	
	m(x) = \frac{m_0}{(1 + \gamma_q x)^2},
\end{equation}
where
$\Psi_q(x, t)=\Psi(x, t)\sqrt{1+\gamma_qx}$
represents a deformation of the wave function solution
$\Psi(x, t)$ and the parameter $\gamma_q \equiv (1-q)/l_0$
controls the variation of the mass in relation to
the position and $l_0$ is a characteristic length.
In terms of the linear deformed derivative
$\mathcal{D}_{q} = (1+\gamma_q x) \partial_x$,
Eq.~(\ref{eq:SE-pdm}) becomes
\cite{CostaFilho-Almeida-Farias-AndradeJr-2011}
\begin{equation}
 \label{eq:SE-pdm-bis}
 \displaystyle
 i\hbar \frac{\partial \Psi_q (x,t)}{\partial t}
		= -\frac{\hbar^2}{2m_0} \mathcal{D}_{q}^2 \Psi_q (x,t)		
         + V(x) \Psi_q (x,t).
\end{equation}
The position-dependent mass, Eq.~(\ref{eq:m(x)}), is the one that allows
(\ref{eq:SE-pdm}) to be rewritten in terms of the deformed derivative $\mathcal{D}_q$
and a constant mass $m_0$, as in (\ref{eq:SE-pdm-bis}).
Parallel to the quantum case, the classical equation of motion
is compactly written by means a deformed Newton's law,
in terms of the dual nonlinear derivative
(\ref{eq:borges-nonlinear-derivative-q-diff}),
as
$m_0 \widetilde{\mathcal{D}}_q^2 x(t) = F(x)$
with
$\widetilde{\mathcal{D}}_q x(t) = \frac{1}{1+\gamma_q x} \frac{dx}{dt}$
(see \cite{Costa-Borges-2018}).
Other possible functions $m(x)$ can lead to different deformed derivatives.
The theoretical approach for the deformed Schr\"odinger equation
(\ref{eq:SE-pdm-bis}) have been applied in the context of
anharmonic potentials
\cite{CostaFilho-Alencar-Skagerstam-AndradeJr-2013,Costa-Borges-2018},
Si and Ge quantum wells \cite{Barbagiovanni-2014},
information theory
\cite{Costa-Gomez-2018,
      Nascimento-Ferreira-Aguiar-Guedes-CostaFilho-2018},
and quasi-periodic potentials \cite{Costa-Gomez-Santos-2020}.

A different generalized nonlinear derivative,
formulated by Nobre {\it et al.} \cite{Nobre-RegoMonteiro-Tsallis-2011},
has also been used to describe position-dependent mass systems.
We represent it by $\widetilde{\mathfrak{D}}_{q}$:
\begin{equation}
 \label{eq:nobre-nonlinear-derivative}
 \widetilde{\mathfrak{D}}_{q} f(u) = [f(u)]^{1-q} \frac{df(u)}{du}.
\end{equation}
It is possible to define its dual linear deformed derivative, as
\begin{equation}
 \label{eq:nobre-linear-derivative}
 \mathfrak{D}_{q} f(u) = \frac{1}{u^{1-q}}\frac{df(u)}{du}.
\end{equation}
The linear and nonlinear deformed derivatives
$\mathcal{\mathcal{D}}_q$ and $\widetilde{\mathfrak{D}}_q$ satisfy
$\mathcal{D}_q \exp_q (u) = \widetilde{\mathfrak{D}}_{q}
                            \exp_q (u) = \exp_q (u)$,
--- Eq.~(\ref{eq:borges-linear-derivative-q-diff}) is
the linear eigenfunction of the $q$-exponential function,
and Eq.~(\ref{eq:nobre-nonlinear-derivative})
is the nonlinear eigenfunction of the $q$-exponential function ---,
and the generalized nonlinear and linear derivatives
(\ref{eq:borges-nonlinear-derivative-q-diff}) and
(\ref{eq:nobre-linear-derivative}) satisfy
$\widetilde{\mathcal{D}}_{q} \ln_q (u) =\mathfrak{D}_{q} \ln_q (u) = 1/u$.
The deformed derivatives
(\ref{eq:nobre-nonlinear-derivative})
and
(\ref{eq:nobre-linear-derivative})
constitute conformable derivative operators in
fractional calculus \cite{Rosa-2018}.

The second derivative of the deformed linear versions
follow the usual derivatives rules:
\begin{eqnarray}
 \label{eq:q-deformed-second-derivative}
 \! \! \!
 \mathcal{D}_q^2 f(u)
 &=& \mathcal{D}_q [\mathcal{D}_q f(u)] \nonumber \\
 &=&[1+(1-q)u]\frac{d}{du} \left\{[1+(1-q)u]\frac{df(u)}{du}\right\}
\end{eqnarray}
and
\begin{eqnarray}
 \label{eq:Nobre-second-derivative}
 \mathfrak{D}_q^2 f(u)
 &=& \mathfrak{D}_q [\mathfrak{D}_q f(u)] \nonumber \\
 &=& \frac{1}{u^{1-q}}\frac{d}{du}
 \left[ \frac{1}{u^{1-q}}\frac{df(u)}{du} \right].
\end{eqnarray}
The second derivative of the deformed nonlinear versions, differently,
must obey the definitions:
\begin{equation}
 \label{eq:q-deformed-second-dual-derivative}
 \widetilde{\mathcal{D}}_q^2 f(u)
 \equiv
 \frac{1}{1+(1-q)f(u)}\frac{d}{du}
 \left[\frac{1}{1+(1-q)f(u)} \frac{df(u)}{du}\right]
\end{equation}
and
\begin{equation}
 \label{eq:Nobre-second-dual-derivative}
 \widetilde{\mathfrak{D}}_q^2 f(u)
 \equiv
  [f(u)]^{1-q} \frac{d}{du} \left\{ [f(u)]^{1-q} \frac{df(u)}{du} \right\},
\end{equation}
i.e.,
$
\widetilde{\mathcal{D}}_q^2 f(u)
\neq \widetilde{\mathcal{D}}_q [\widetilde{\mathcal{D}}_q f(u)]
$
and
$
\widetilde{\mathfrak{D}}_q^2 f(u)
\neq \widetilde{\mathfrak{D}}_q [\widetilde{\mathfrak{D}}_q f(u)].
$
Higher order derivatives are found analogously.

The nonlinear deformed derivative (\ref{eq:nobre-nonlinear-derivative})
can be used to formulate the nonlinear Fokker-Planck equation
proposed in \cite{Plastino-Plastino-1995}.
The deformed PDF satisfies the equation
\begin{equation}
 \label{eq:nonlinear-FPE}
 \frac{\partial P_q(x,t)}{\partial t}
      = - \frac{\partial}{\partial x}  [A(x) P_q( x,t)]
        + \frac{\Gamma}{2}\frac{\partial^2}{\partial x^2}
          \left[ P_q(x,t) \right]^{2-q},
\end{equation}
which is equivalent to
\begin{equation}
 \label{eq:nonlinear-FPE-and-derivative}
 \widetilde{\mathfrak{D}}_{q,t} P_q(x,t)
  = - \widetilde{\mathfrak{D}}_{q,x} [A(x) P_q(x,t)]
    + \frac{\Gamma_q}{2} \widetilde{\mathfrak{D}}_{q,x}^2 P_q(x,t),
\end{equation}
where $\Gamma_q = (2-q)\Gamma$, and $q<2$,
as it is considered in Ref.~\cite{Plastino-Plastino-1995}.
The use of a linear confining potential $A(x)$
leads to a $q$-Gaussian distribution
\cite{Plastino-Plastino-1995,Tsallis-Bukmann-1996,Borland-PRE-1998}.
The nonlinear Fokker-Planck equation (\ref{eq:nonlinear-FPE-and-derivative})
has been useful in the description of some experiments, \emph{e.g.,}:
single ions in radio frequency traps interacting with a classical buffer gas
\cite{DeVoe-2009},
momentum distribution of cold atoms in dissipative optical lattices
\cite{Lutz}.

By last, a nonlinear generalization of the Schr\"odinger equation,
i.e., the $q$-deformed nonlinear version
given by \cite{Nobre-RegoMonteiro-Tsallis-2011}
\begin{equation}
\label{eq:SE-nobre}
\frac{\partial \Phi_q (x, t)}{\partial t}
     = -\frac{1}{2-q}\frac{\hbar^2}{2m_0}
       \frac{\partial^2 \Phi_q^{2-q} (x,t)}{\partial x^2} 		
       + V(x) {\Phi}_q^q (x,t),
\end{equation}
where $\Phi_q (x, t)$ is a deformation of the
solution $\Phi (x, t)$ (corresponding to $q\rightarrow1$),
can be recast by means of the dual (nonlinear) derivative
$\widetilde{\mathfrak{D}}_{q}$ as
\begin{equation}
 \label{eq:SE-nobre-bis}
 i\hbar \widetilde{\mathfrak{D}}_{q,t} \Phi_q (x,t)
      = -\frac{\hbar^2}{2m_0}\widetilde{\mathfrak{D}}_{q,x}^2 \Phi_q (x,t)		
        + V(x) \Phi_q (x,t),
\end{equation}
with $\widetilde{\mathfrak{D}}_{q,t}$ and $\widetilde{\mathfrak{D}}_{q,x}$
standing for the nonlinear derivatives with respect to time and position
variables $t$ and $x$, respectively.
The nonlinear Schr\"odinger equation (\ref{eq:SE-nobre}) has attracted
the attention of theoretical physicists due to some of its features.
In particular, solutions of (\ref{eq:SE-nobre}) have a solitary-wave behavior
(see, for instance, \cite{RegoMonteiro-PLA-2020}
 and references therein),
a typical phenomenon in several areas of physics,
such as nonlinear optics, plasma physics and superconductivity.

\subsection{Linear and nonlinear deformed operators}

Motivated by the deformed derivatives
$\mathcal{D}_q$ and $\mathfrak{D}_q$ and their duals,
we define more general operators for an arbitrary deformation $h(u)$.
Without loss of generality, $h(u)$ is assumed to be
an infinitely differentiable function.
In this context, a linear deformed derivative is defined as
\begin{equation}
 \label{eq:generalized-linear-derivative}
 \mathcal{D}_{[h]} f(u) = \frac{1}{h(u)} \frac{df}{du}
\end{equation}
and its dual nonlinear derivative,
\begin{equation}
 \label{eq:generalized-nonlinear-derivative}
 \widetilde{\mathcal{D}}_{[h]} f(u) = h(f) \frac{df(u)}{du}.
\end{equation}
The function $h(u)$ specifies the deformation.
The infinitesimal element
$ du_{[h]} \equiv h(u)du $
implies
$ u_{[h]}(u) = \int^{u} h(u')du', $
that may be considered as a deformed independent variable,
leading to the equivalence
\begin{equation}
 \label{eq:du_h=d_hu}
 d u_{[h]} = d_{[h]} u,
\end{equation}
i.e., the differential of the deformed variable $u_{[h]}$
is equal to the deformed differential of the ordinary variable $u$.
The deformed derivative operator $\mathcal{D}_{[h]} f(u)$
is regarded as the rate of variation of the function $f(u)$
with respect to the variation of the deformed variable
$u_{[h]}$, or, equivalently,
the rate of variation of the function $f(u)$ with respect to a
deformed variation of the variable $u$, denoted by $d_{[h]}u$:
\begin{equation}
 \label{eq:generalized-linear-derivative-2}
 \mathcal{D}_{[h]} f(u) = \frac{df(u)}{d u_{[h]}}
                        = \frac{df(u)}{d_{[h]} u}.
\end{equation}
This operator is linear regarding the dependent variable $f$.
Analogously, the deformed derivative operator
$\widetilde{\mathcal{D}}_{[h]} f(u)$
may be viewed as the rate of a generalized variation of the function $f(u)$
with respect to the ordinary variation of the independent variable $u$,
and Eq.~(\ref{eq:generalized-nonlinear-derivative}) becomes
\begin{equation}
\label{eq:generalized-nonlinear-derivative-2}
 \widetilde{\mathcal{D}}_{[h]} f(u) = \frac{d_{[h]} f(u)}{d u}.
\end{equation}
This derivative is nonlinear regarding the dependent variable $f$.
It is straightforwardly verified that
$\mathcal{D}_{[h]} f = 1/\widetilde{\mathcal{D}}_{[h]} f^{-1}$,
expressing the duality between them.
Thus, the $q$-deformed derivatives $\mathcal{D}_q$
and $\mathfrak{D}_q$ (Eqs.~(\ref{eq:borges-linear-derivative-q-diff})
and (\ref{eq:nobre-linear-derivative}))
are obtained as special cases
of (\ref{eq:generalized-nonlinear-derivative-2})
for $h(u)=\frac{1}{1+(1-q)u}$ and $h(u)=u^{1-q}$
respectively.
Similarly, $\widetilde{\mathcal{D}}_q$ and $\widetilde{\mathfrak{D}}_q$
(Eqs.~(\ref{eq:borges-nonlinear-derivative-q-diff}) and
(\ref{eq:nobre-nonlinear-derivative})) are special cases of
Eq.~(\ref{eq:generalized-nonlinear-derivative-2})
for $h(f) = \frac{1}{1+(1-q)f}$ and $h(f) = f^{1-q}$, respectively.

Second (and higher) derivatives of these generalized operators follow
the same corresponding rules,
\begin{equation}
\mathcal{D}_{[h]}^2 f(u) = \frac{1}{h(u)} \frac{d}{du}
\left[ \frac{1}{h(u)} \frac{d f(u)}{du} \right]
\end{equation}
and
\begin{equation}
\widetilde{\mathcal{D}}_{[h]}^2 f(u) = h(f) \frac{d}{du}
\left[ h(f) \frac{d f(u)}{du} \right],
\end{equation}
i.e.,
$\mathcal{D}_{[h]}^2 f(u) = \mathcal{D}_{[h]} [\mathcal{D}_{[h]}f(u)]$,
but
$\widetilde{\mathcal{D}}_{[h]}^2 f(u)\neq
 \widetilde{\mathcal{D}}_{[h]}[\widetilde{\mathcal{D}}_{[h]}f(u)]$.

\section{\label{sec:FP-inhomogeneous-media}
         Diffusion processes in inhomogeneous
         media from position-dependent mass}

We revisit the path outlined in Section \ref{sec:linear-nonlinear-FPE}
provided with a position-dependent effective mass $m(x)$
in an inhomogeneous medium, and express the FPE as a homogeneous one
by means of the deformed derivative associated with the $q$-algebra.
We provide a version of the $H$-theorem along with a discussion about
the nonlinear FPE within the context of the deformed derivative.

\subsection{\label{deduction-deformed-FPE}
            Nonlinear Langevin equation}

The nonlinear Langevin equation is (\cite{vanKampen-book}, equation (4.3))
\begin{equation}
 \label{eq:nonlinear-Langevin}
 \dot{x}=A(x)+L(t)
\end{equation}
with $A(x)$ being a generic force and $L(t) = \xi (t)/\lambda_0$
the unpredictable term whose stochastic properties are
$\langle L(t) \rangle=0$ and
$\langle L(t)L(t^{\prime})\rangle=\Gamma\delta(t-t^{\prime})$
(fast variation due to individual molecule collisions).
The fully nonlinear Langevin equation
\begin{equation}
 \label{eq:fullynonlinear-Langevin}
 \dot{x}=A(x)+C(x)L(t)
\end{equation}
results equivalent to (\ref{eq:nonlinear-Langevin})
by means of the change of variable
\begin{equation}
 \overline{x}=\int \frac{dx}{C(x)}, \quad
 \frac{A(x)}{C(x)}=\overline{A}(\overline{x}), \quad
 \overline{P}(\overline{x})=P(x)C(x)
\end{equation}
with the coefficient
$C(x)$ representing the heterogeneities of the medium.
Thus, the nonlinear Langevin equations
(\ref{eq:nonlinear-Langevin}) and (\ref{eq:fullynonlinear-Langevin})
have their equivalent Fokker-Planck equations,
given by (equations (4.7) and (4.8) of \cite{vanKampen-book})
\begin{subequations}
\begin{eqnarray}
\frac{\partial \overline{P}(\overline{x},t)}{\partial t}&=&
-\frac{\partial}{\partial x}[\overline{A}(\overline{x})\overline{P}(\overline{x},t)]+
\frac{\Gamma}{2}\frac{\partial^2\overline{P} (\overline{x},t)}{\partial \overline{x}^2} \\
\frac{\partial P(x,t)}{\partial t}&=& \label{eq:fullynonlinear-FPE}
-\frac{\partial}{\partial x}[A(x)P (x,t)]
\nonumber \\
&&
+\frac{\Gamma}{2}\left\{
\frac{\partial}{\partial x}C(x)\left[ \frac{\partial}{\partial x}C(x)P (x,t) \right]
\right\}.
\end{eqnarray}
\end{subequations}
The coefficient $C(x)$ can be identified with the variable temperature $T(x)$,
if $T(x)/\mu(x)$ is constant, according to the van Kampen diffusion FPE
\eqref{eq:van-Kampen-equation}.

\subsection{Generalized Fokker-Planck equation for inhomogeneous media}

The Lagrangian for a position-dependent mass system is
\begin{equation}
 \label{eq:lagrangian-pdm}
 \mathcal{L}(x, \dot{x}, t) = \frac{1}{2}m(x) \dot{x}^2 - U(x, t).
\end{equation}
The Langevin equation for a position-dependent damping coefficient $\lambda(x)$
and
$Q = \frac{1}{2}m(x) \lambda(x) \dot{x}^2$
follows from the Euler-Lagrange equation (\ref{eq:Euler-Lagrange}):
\begin{equation}
 \label{eq:masslangevin-pdm}
 m(x)\ddot{x} + \frac{1}{2}m'(x)\dot{x}^2
 = - m(x){\lambda}(x)\dot{x} + F(x) + R(t)
\end{equation}
where now we have a new kinetic term $\frac{1}{2}m'(x)\dot{x}^2$
due to the position-dependent mass $m(x)$.
We see the standard Langevin equation (\ref{eq:Langevin-equation})
follows for $m(x)=m_0$ and $\lambda(x)=\lambda_0$.
In the overdamped limit ($\lambda(x) \gg \tau^{-1}$)
the left-hand side of (\ref{eq:masslangevin-pdm}) vanishes, so we obtain
\begin{equation}\label{eq:masslangevin-pdm-2}
 \frac{dx}{dt}
= \frac{1}{m(x) {\lambda}(x)}[F(x)+R(t)]
\end{equation}
or, alternatively,
\begin{equation}
\label{eq:deformed-Langevin}
  \widetilde{\mathcal{D}}_{[\kappa]} x(t)
	= \frac{1}{\lambda_0}[f(x)+{\xi}(t)],
\end{equation}
a deformed Langevin equation with the dimensionless deformation parameter
$\kappa(x)=\frac{\lambda(x)}{\lambda_0}\frac{m(x)}{m_0}$.

Inhomogeneous diffusion can be alternatively described
by the Caldeira-Leggett's model \cite{Caldeira-Leggett-1983}
in terms of
a system interacting with an inhomogeneous environment composed by a large
number $N$ of harmonic oscillators having equilibrium positions $\mathcal{Q}_n(x,t)$.
For the special case
$\mathcal{Q}_n(x,t)=\mathcal{Q}(x)$ for all $n=1,\ldots,N$, the overdamped Caldeira-Leggett's Langevin equation
(equation (27) of \cite{Illuminati-94})
and the overdamped PDM Langevin equation \eqref{eq:masslangevin-pdm-2}
(or equivalently \eqref{eq:deformed-Langevin}) are the same subjected to
the conditions $-V'(x)/\mathcal{Q}'(x)^2=F(x)/\kappa(x)$ and
$\kappa(x) = \mathcal{Q}'(x)$ with $-V'(x)/\mathcal{Q}'(x)$ an effective force,
thus showing a connection between the Caldeira-Leggett's model
and the position-dependent mass systems.

The FPE for an inhomogeneous medium of mass $m(x)$
and dumping coefficient $\lambda(x)$
follows from Eq.~(\ref{eq:generalized-difusion}):
\begin{eqnarray}
 \label{eq:fpe-inhomogeneous-1}
 \frac{\partial P}{\partial t}
 &=&
     - \frac{\partial}{\partial x}
       \left[ \frac{A(x)}{\kappa (x)} P(x,t) \right]
 \nonumber \\
 & & + \frac{D_0}{\lambda_0^2}
       \frac{\partial}{\partial x}
       \left\{
        \left[
              - \frac{\kappa'(x)}{\kappa^3 (x)}
              + \frac{1}{{\kappa^2 (x)}} \frac{\partial}{\partial x}
        \right] P(x,t)
       \right\}
 \nonumber \\
&=&
    - \frac{\partial }{\partial x}
      \left[ \frac{A(x)}{{\kappa}(x)}  P(x,t) \right]
 \nonumber \\
 & & + \frac{\Gamma}{2}
       \frac{\partial}{\partial x}
       \left\{
              \frac{1}{{\kappa}(x)}
              \frac{\partial}{\partial x}
              \left[
                     \frac{1}{{\kappa}(x)} P(x,t)
              \right]
       \right\}.
\end{eqnarray}
The van Kampen's FPE \eqref{eq:van-Kampen-equation}
and the inhomogeneous
FPE \eqref{eq:fpe-inhomogeneous-1} differ from each other
if the temperature and mobility are not
inversely proportional to the deformed parameter
$\kappa(x)$ (we address to this point later, see equation \eqref{eq:choice}).
If we define
\begin{equation}
 \label{eq:PDM-diffusion-coefficient}
 D(x) \equiv D_0/\kappa^2(x) \geq 0
\end{equation}
as the position-dependent diffusion coefficient,
then we can recast the Fokker-Planck equation for an inhomogeneous medium
(\ref{eq:fpe-inhomogeneous-1}) as
\begin{eqnarray}
 \label{eq:fpe-inhomogeneous-2}
\! \! \! \! \!
 \frac{\partial P}{\partial t}
 &=&
     - \frac{\partial }{\partial x}
        \left[ \sqrt{\frac{D(x)}{D_0}} A(x) P(x,t) \right]
 \nonumber \\
 & & + \frac{D_0}{\lambda_0^2}
       \frac{\partial}{\partial x}
        \left\{
               \sqrt{\frac{D(x)}{D_0}}
               \frac{\partial}{\partial x}
               \left[
				\sqrt{\frac{D(x)}{D_0}} P(x,t)
               \right]
        \right\}.
\end{eqnarray}
Eq.~(\ref{eq:fpe-inhomogeneous-1})
(or (\ref{eq:fpe-inhomogeneous-2}))
obeys the probability conservation law
$\partial P /\partial t = -\partial J/\partial x$
with the current of probability
\begin{eqnarray}
\label{eq:current-probability}
 J(x,t) &=& E(x) P(x,t)
		 - \frac{\Gamma}{2}
		\frac{1}{\kappa^2 (x)}\frac{\partial P}{\partial x}
		 \nonumber \\
		 &=& E(x) P(x,t)	
 		 - \frac{D(x)}{\lambda_0^2} \frac{\partial P}{\partial x},
\end{eqnarray}
and the drift coefficient
\begin{eqnarray}
\label{eq:drift-coefficient}
E(x) &=& \frac{A(x)}{\kappa(x)}-\frac{\Gamma}{2}\frac{\kappa'(x)}{\kappa^3(x)}
		\nonumber \\
  	 &=& \sqrt{\frac{D(x)}{D_0}}A(x)+\frac{D'(x)}{2\lambda_0^2},
\end{eqnarray}
where the first term is associated with the confining potential
and the second term is proportional to the derivative
of the diffusion coefficient.
Notice that $D(x)$ has contributions of the viscosity
and of the mass of the particles,
where the latter may be position-dependent
due to the non-isotropy of the space.
Other formulations for the FPE in inhomogeneous media have been reported.
For instance in Ref.\ \cite{Sicuro-Rapcan-Tsallis-2016}
a current of probability (\ref{eq:current-probability}),
whose second term depends on the power law of the PDF, has been considered.
In the present work, we restrict our analysis to linear current densities
in inhomogeneous media.

The stationary solution for reflecting boundary conditions
($\lim_{x \rightarrow \pm \infty} J(x,t) = 0$)
may be expressed by the integral form
\begin{align}\label{eq:deformed-stationary-solutions}
 P^{(\textrm{st})}(x)
 &=
   C \kappa(x)
   \exp \left[
               \frac{2}{\Gamma} \int^x A(x') \kappa(x' )d x'
        \right]
\nonumber \\
&=
\frac{C}{\sqrt{D(x)/D_0}}
\exp \left[
      \frac{2}{\Gamma} \int^x \frac{A(x')}{\sqrt{D(x')/D_0}} d x'
        \right].
\end{align}
The FPE for an inhomogeneous medium (\ref{eq:fpe-inhomogeneous-1})
can be formally rewritten for a \textit{homogeneous} medium,
and the inhomogeneity is encompassed by an appropriate deformation
of the derivative, according to Eq.~(\ref{eq:generalized-linear-derivative})
(written as a partial derivative)
with $h(x) \equiv \kappa(x)$, and the transformation
\begin{equation}
\label{eq:P_kappa}
\mathcal{P}_{[\kappa]}(x, t) = \frac{P(x, t)}{\kappa (x)},
\end{equation}
so,
\begin{equation}
 \label{eq:fpe-inhomogeneous-deformed-derivative}
 \frac{\partial \mathcal{P}_{[\kappa]} (x, t)}{\partial t}
 = - \mathcal{D}_{[\kappa]} \left[
     A(x) \mathcal{P}_{[\kappa]} (x,t)
     \right]
   + \frac{\Gamma}{2} \mathcal{D}_{[\kappa]}^2 \mathcal{P}_{[\kappa]} (x,t).
\end{equation}
The deformed PDF $\mathcal{P}_{[\kappa]}(x, t)$ satisfies a generalized version
of the normalization condition,
\begin{equation}
 \label{eq:q-norm}
 \int \mathcal{P}_{[\kappa]}(x, t) d_{[\kappa]}x = 1.
\end{equation}
As a consequence of $ d x_{[\kappa]} = d_{[\kappa]} x$
(Eq.~(\ref{eq:du_h=d_hu}) with $h=\kappa$),
$\mathcal{P}_{[\kappa]}(x, t)$ is normalized
in the deformed space $x_{[\kappa]}$.
The stationary solution of
Eq.~(\ref{eq:fpe-inhomogeneous-deformed-derivative}) is
\begin{equation}\label{eq:stationary-solution-deformed-space}
 \mathcal{P}_{[\kappa]}^{(\textrm{st})}(x)
  = C \exp \left[
                  \frac{2}{\Gamma} \int^x A(x') d_{[\kappa]} x'
           \right],
\end{equation}
which is entirely written as an integral
with a deformed differential $d_{[\kappa]} x'$.
In the limit of absence of deformation ($\kappa\rightarrow1$)
we recover the standard stationary solution
(\ref{eq:stationary-sem-deformacao}).

It shall be imposed $\kappa(x) = \sqrt{\frac{m(x)}{m_0}}$
on Eq.~(\ref{eq:fpe-inhomogeneous-deformed-derivative}), thereby
\begin{subequations}
 \begin{eqnarray}
 \label{eq:dumping-diffusion}
 &\lambda(x) = \displaystyle
 \frac{\lambda_0}{\kappa(x)}, \\
 &D(x)= \displaystyle
 D_0 \left(\frac{\lambda(x)}{\lambda_0}\right)^{2},
 \end{eqnarray}
\end{subequations}
i.e., the deformation of the space $\kappa(x)$
(or equivalently, the mass $m(x)$)
univocally determines the dumping and diffusion coefficients
that are compatible with the deformed linear Fokker-Planck equation
(\ref{eq:fpe-inhomogeneous-deformed-derivative}).

The equivalence between the position-dependent Langevin equation
(\ref{eq:masslangevin-pdm-2})
and the fully nonlinear one (\ref{eq:fullynonlinear-Langevin})
is established by identifying
$C(x)$ with $1/\kappa(x)$, $A(x)$ with $f(x)/(\lambda_0\kappa(x))$ and
$L(t)$ with $\xi(t)/\lambda_0$ and recalling that $F(x)=m_0f(x)$ and
$R(t)=m_0\xi(t)$. Moreover,
$C(x)\frac{\partial}{\partial x}=\mathcal{D}_{[\kappa]}$ and
$C(x)P=\mathcal{P}_{\kappa}$ and multiplying both sides of
(\ref{eq:fullynonlinear-FPE}) by $C(x)$ the equivalence between
the deformed FPE (\ref{eq:fpe-inhomogeneous-deformed-derivative})
and (\ref{eq:fullynonlinear-FPE}) also follows.

\subsection{\label{subsec:deformed-theorem-H}
            $H$-Theorem for inhomogeneous FPE: entropy of the medium}

Applying the strategy employed in \cite{Schwammle-Curado-Nobre-2007}
for the generalized free energy functional
\begin{equation}
 \label{eq:q-helmholtz}
 \mathcal{F}[\mathcal{P}_{[\kappa]}]
 =
 \int \Phi_{[\kappa]} [x, \mathcal{P}_{[\kappa]}(x, t)] d_{[\kappa]} x
 = \mathcal{U} - \theta \mathcal{S},
\end{equation}
(with $\theta$ an inverse of the Lagrange multiplier),
it is possible to establish a generalized version
of the $H$-theorem for the inhomogeneous FPE
(\ref{eq:fpe-inhomogeneous-deformed-derivative}).
The first term is
\begin{equation}
 \label{eq:q-energy}
 \mathcal{U}[\mathcal{P}_{[\kappa]}]
 = \int \vartheta_{[\kappa]} (x) \mathcal{P}_{[\kappa]} (x, t) d_{[\kappa]} x,
\end{equation}
where $\vartheta_{[\kappa]}(x)$ corresponds to an auxiliary potential,
while the second term of (\ref{eq:q-helmholtz}) is a deformed entropy functional
\begin{equation}
 \label{eq:q-entropy}
 \mathcal{S}[\mathcal{P}_{[\kappa]}] =
 \int s_{[\kappa]}[\mathcal{P}_{[\kappa]}(x, t)] d_{[\kappa]} x,
\end{equation}
with the usual convex conditions
$s_{[\kappa]}[0] = s_{[\kappa]}[1] = 0$
and  $d^2s_{[\kappa]}/d\mathcal{P}_{[\kappa]}^2 \leq 0$.
The time derivative of (\ref{eq:q-helmholtz}) is
\begin{eqnarray}
	\frac{d\mathcal{F}}{dt}
	&=&
	\int
	\left[\vartheta_{[\kappa]}(x) -\theta \frac{ds_{[\kappa]}}{d\mathcal{P}_{[\kappa]}} \right]
	\frac{\partial \mathcal{P}_{[\kappa]}}{\partial t} d_{[\kappa]} x
	\nonumber \\
	&=&
	\int \left[
		\vartheta_{[{\kappa}]}(x) -\theta \frac{d s_{[\kappa]}}{d\mathcal{P}_{[\kappa]}}
	\right]
	\nonumber \\
	&& \qquad \times \mathcal{D}_{[\kappa]} \left[ - A(x) \mathcal{P}_{[\kappa]}
	+ \frac{\Gamma}{2} \mathcal{D}_{[\kappa]} \mathcal{P}_{[\kappa]}
	\right] d_{[\kappa]} x
	\nonumber \\
	&=&
	- \int  \mathcal{P}_{[\kappa]}
	\left[ \mathcal{D}_{[\kappa]} \vartheta_{[{\kappa}]}(x)
	-\theta \frac{d^2 s_{[\kappa]}}{d\mathcal{P}_{[\kappa]}^2}
	\mathcal{D}_{[\kappa]} \mathcal{P}_{[\kappa]}\right]
	\nonumber \\
	&& \qquad \times \left[ - A(x)
	+ \frac{\Gamma}{2} \frac{ \mathcal{D}_{[\kappa]} \mathcal{P}_{[\kappa]}}{\mathcal{P}_{[\kappa]}}
	\right] d_{[\kappa]} x.
\end{eqnarray}
The definition of $\theta =\frac{\Gamma}{2},$
$\mathcal{D}_{[\kappa]} \vartheta_{[\kappa]}(x) = -A(x)$,
and
$d^2 s_{[\kappa]}/d\mathcal{P}_{[\kappa]}^2  = -1/\mathcal{P}_{[\kappa]}$
imply
\begin{equation}
 \frac{d\mathcal{F}}{dt} \leq 0, \qquad \forall t \geq 0.
\end{equation}
The Boltzmann-Gibbs entropy density for a deformed probability space is
$s_{[\kappa]}[\mathcal{P}_{[\kappa]}]  =
 -\mathcal{P}_{[\kappa]} \ln \mathcal{P}_{[\kappa]}$,
with $\vartheta_{[\kappa]}(x) = - \int A(x) d_{[\kappa]} x$.
The general entropy of the system is given by the integral
\begin{equation}
\label{eq:total-entropy}
 \mathcal{S}
 = - \int
    \mathcal{P}_{[\kappa]} (x,t)
    \ln \mathcal{P}_{[\kappa]} (x,t) d_{[\kappa]} x.
\end{equation}
The meaning of $\mathcal{S}$ can be examined by
transforming back the variables with
(\ref{eq:P_kappa}):
\begin{eqnarray}
 \label{eq:deformed-entropy}
 \mathcal{S} &=& -\int P(x,t) \ln [P(x,t)/\kappa (x)] dx
 \nonumber \\
             &=& S_{\text{BG}} + \langle \ln [\kappa(x)] \rangle.
\end{eqnarray}
The general entropy $\mathcal{S}$ for an inhomogeneous medium
is the sum of two terms: the Boltzmann-Gibbs entropy
$S_{\text{BG}} = -\int P(x,t) \ln P(x,t) dx$
associated with the distribution of particles,
and a residual contribution resulting from the inhomogeneity of the medium
$S_{\text{medium}} = \int P(x,t) \ln [\kappa(x)]dx $.
The quantity $\mathcal{S}$ in (\ref{eq:deformed-entropy}) looks like
the Kullback-Leibler divergence\footnote{Only if $\kappa(x)$ is normalized.},
or relative entropy
\cite{Kullback-Leibler-1951}
$S_{\text{KL}}(P,P_0)= -\int P(x,t)\ln\left[P(x, t)/P_0(x,t)\right]dx$,
with the reference distribution
$P_0(x,t)$ replaced by $\kappa(x)$.

\section{\label{sec:solutions}$q$-Deformed Fokker-Planck equation}

In this Section we focus on a particular generalization
of the FPE associated to the $q$-derivative
given by Eq.~(\ref{eq:borges-linear-derivative-q-diff}).
As previously mentioned, the $q$-derivative originates
from the $q$-sum of the $q$-algebra
$a\oplus b=a+b+(1-q)ab$
and it is related to the quantum approach of
the generalized displacement
operator \cite{CostaFilho-Almeida-Farias-AndradeJr-2011,CostaFilho-Alencar-Skagerstam-AndradeJr-2013}
$\mathcal{T}_q(a)|x\rangle=|x+a+\gamma_q xa\rangle$ ($\gamma_q=(1-q)/\xi$,
$\xi$ a characteristic length), thus giving place to
a $q$-deformed Schr\"odinger equation in terms of the linear $q$-derivative
\eqref{eq:borges-nonlinear-derivative-q-diff}
$\mathcal{D}_q$
and with a position-dependent mass inherited by the $q$-algebra.
As we shall see in Sub-section VI B, the linear
$q$-deformed derivative provides a simple choice for the $q$-deformation, or equivalently for the
variable temperature profile in the van Kampen's sense, that allows to obtain
the inverse Gamma distribution for the superstatistical probability density $f(\beta)$
in the overdamped limit, employed in \cite{vanderStraeten} to model wind velocity fluctuations.
For this purpose, we consider a medium with a diffusion coefficient
depending on the position $x$ of the form
\begin{equation}
 \label{eq:D(x)}
 D(x) = D_0(1+\gamma_q x)^2
\end{equation}
which corresponds, according to Eq.~(\ref{eq:dumping-diffusion}),
to a deformation
\begin{equation}\label{eq:q-deformation}
\kappa(x)=\frac{1}{1+\gamma_qx}
\end{equation}
and a dumping coefficient $\lambda(x)=\lambda_0(1+\gamma_q x)$.
Using the deformed PDF (\ref{eq:P_kappa}) and the $q$-derivative
(\ref{eq:borges-linear-derivative-q-diff}),
the $q$-deformed Fokker-Planck equation
(\ref{eq:fpe-inhomogeneous-deformed-derivative})
can be recast as
\begin{equation}
\label{eq:q-FPE}
\frac{\partial \mathcal{P}_q(x, t)}{\partial t} =
	-\mathcal{D}_{q}[A(x) \mathcal{P}_q(x,t)]
 	+
	\frac{\Gamma}{2}
	\mathcal{D}_{q}^2 \mathcal{P}_q(x,t)
\end{equation}	
provided the deformed normalization condition
$ \int \mathcal{P}_q(x,t)d_q x =1.  $
It is clear that the $q$-deformed FPE is simply the standard one
(compare with Eq.~(\ref{eq:fokker-planck}))
but replacing the usual derivative $d/dx$ and the PDF $P(x,t)$
by their $q$-deformed versions
$\mathcal{D}_{q}$ and $\mathcal{P}_q(x,t)$.
This remark indicates that, when the diffusion coefficient
depends on the position, it is possible to express
the inhomogeneous FPE as the standard one
having a constant diffusion coefficient,
with the inhomogeneity contained in the deformed derivatives.
In the next Sub-Sections we analyze the effect of the deformation
on the solutions of Eq.~(\ref{eq:q-FPE}) and their physical consequences
on the diffusion processes.

\subsection{\label{stationary}
            Stationary solution}

In order to obtain the stationary solution of the $q$-deformed FPE
(\ref{eq:q-FPE}), we rewrite it as
\begin{equation}
 \frac{\partial \mathcal{P}_q}{\partial t} =
 - \mathcal{D}_{q} \mathcal{J}_q(x, t),
\end{equation}
where $\mathcal{J}_q(x, t)$ is a $q$-deformed probability current density
\begin{equation}
	 \mathcal{J}_q(x, t) = A(x) \mathcal{P}_q(x, t)
	-\frac{\Gamma}{2}
	  \mathcal{D}_{q} \mathcal{P}_q(x, t).
\end{equation}
Using the $q$-deformed integral (see \cite{Borges-2004}),
\begin{equation}
 \frac{\partial}{\partial t} \int_{x_i}^{x_f} \mathcal{P}_q(x,t) d_q x
 = \mathcal{J}_q(x_f, t) - \mathcal{J}_q(x_i, t).
\end{equation}
According to the deformed $q$-normalization condition,
the conservation of the total deformed probability is guaranteed only if
$\mathcal{J}_q(x_f, t) = \mathcal{J}_q(x_i, t)$.
The stationary solution
($\partial \mathcal{P}_q^{(\text{st})}/\partial t = 0 $),
with reflecting boundary conditions
($\mathcal{J}_q(x, t) = 0, \ \forall x$), satisfies
\begin{equation}
	A(x) \mathcal{P}_q^{(\text{st})}(x) =
	\frac{\Gamma}{2}
\mathcal{D}_{q} \mathcal{P}_q^{(\text{st})} (x),
\end{equation}
leading to
\begin{equation}
 \label{eq:P_q(x)-stationary}
 \mathcal{P}_q^{(\text{st})}(x)
 = C_q \exp \left[
                   \frac{2}{\Gamma} \int_{0}^{x} A(x') d_q x'
            \right],
\end{equation}
where $C_q$ is a normalization constant.

\subsection{\label{general-solution}
            General solution}

The general solution for the $q$-deformed FPE can be obtained by
the method of separation of variables \cite{Araujo-2012}.
In this direction, let us consider the $q$-deformed FPE (\ref{eq:q-FPE})
expressed as
\begin{equation}
 \frac{\partial \mathcal{P}_q(x, t)}{\partial t}
 = \hat{\mathcal{L}}_q \mathcal{P}_q(x, t)
\end{equation}
where $\hat{\mathcal{L}}_q$ is a deformed Fokker-Planck operator
whose action over a function $f(x)$ is
$
 \hat{\mathcal{L}}_q f(x)
 = - \mathcal{D}_{q}[A(x) f(x)]
   + \frac{1}{2}\Gamma \mathcal{D}_{q}^2 f(x).
$
The general solution of the $q$-deformed FPE
can be expanded in a power series of the eigenfunctions $\phi_{q,n}(x)$
with the coefficients $c_n$
and the eigenvalues $\Lambda_n$ of $\hat{\mathcal{L}}_q$, i.e.,
\begin{equation}
\label{eq:expansion}
 \mathcal{P}_q(x, t) = \sum_{n}c_n \phi_{q,n}(x) e^{t\Lambda_n t}.
\end{equation}
By the boundary conditions in $x = x_i$ and $x = x_f$, it follows
\begin{equation}
 \label{eq:conditions-phi}
 \int_{x_i}^{x_f} \hat{\mathcal{L}}_q \phi_{q,n}(x) d_q x
 = \Lambda_n \int_{x_i}^{x_f} \phi_{q,n}(x) d_q x = 0,
\end{equation}
with $x_i,x_f\in(-1/\gamma_q,\infty)$.
Next step is to employ an associated Schr\"odinger equation
for obtaining the explicit formula of the general
solution of the $q$-deformed FPE.
To accomplish this, we use the operator $\hat{\mathcal{K}}_q$ defined by
\begin{equation}
 \hat{\mathcal{K}}_q \psi_{q,n}(x)
 = \frac{\hat{\mathcal{L}}_q [\psi_{q,0}(x) \phi_{q,n}(x)]}{\psi_{q,0}(x)},
\end{equation}
where $\psi_{q,n}(x) = \phi_{q,n}(x)/\psi_{q,0}(x)$
with $\psi_{q,0}(x) = \sqrt{\mathcal{P}_q^{(\textrm{st})}(x)}$.
It is straightforward to show that
$\hat{\mathcal{K}}_q \psi_{q,n} (x) = \Lambda_n \psi_{q,n}(x)$,
and then, from the relation
$\mathcal{D}_{q}[\ln \psi_{q,0}(x)] = A(x)/\Gamma$
and using the operator $\hat{\mathcal{L}}_q$, we obtain
\begin{equation}
 \label{eq:K_q-psi_q}
 \hat{\mathcal{K}}_q \psi_{q}(x) =
 \frac{\Gamma}{2} \mathcal{D}_{q}^2 \psi_{q}(x)
 - \frac{1}{2}
   \left\{
          \frac{1}{\Gamma}[A(x)]^2 + \mathcal{D}_{q}A(x)
   \right\} \psi_{q}(x).
\end{equation}
The operator ($-\hat{\mathcal{K}}_q$)
is the $q$-deformed Hamiltonian operator
\cite{CostaFilho-Almeida-Farias-AndradeJr-2011}
\begin{equation}
\label{q-deformed-Hamiltonian}
	\hat{H}_q = -\frac{\hbar^2}{2m_0}\mathcal{D}_{q}^2
				+ V_{\textrm{ef}}(\hat{x}),
\end{equation}
which is associated to a quantum system having a position-dependent mass
given by Eq.~(\ref{eq:m(x)}) and subjected to an effective potential
of the form
\begin{equation}
 \label{eq:effective-potential}
 V_{\textrm{ef}}(x) = \frac{1}{2}
                      \left\{
                             \frac{1}{\Gamma} [A(x)]^2 + \mathcal{D}_{q}A(x)
                      \right\}.
\end{equation}
Hence, by comparison with the solutions of (\ref{q-deformed-Hamiltonian}),
the general solution of the $q$-deformed FPE results
\begin{equation}
 \label{eq:general-solution-psi}
 \mathcal{P}_q(x, t) = \psi_{q,0}(x) \sum_{n}c_n \psi_{q,n}(x) e^{t\Lambda_n}.
\end{equation}

\section{\label{sec:applications}
         Applications of the $q$-deformed Fokker-Planck equation}

We illustrate the $q$-deformed FPE with two examples of potentials:
the infinite square well
and the confining potential with linear drift coefficient.

\subsection{\label{subsec:infinite-well}
            Infinite square well potential}

\begin{figure*}[!hbt]
\centering
\begin{minipage}[h]{0.32\linewidth}
\includegraphics[width=\linewidth]{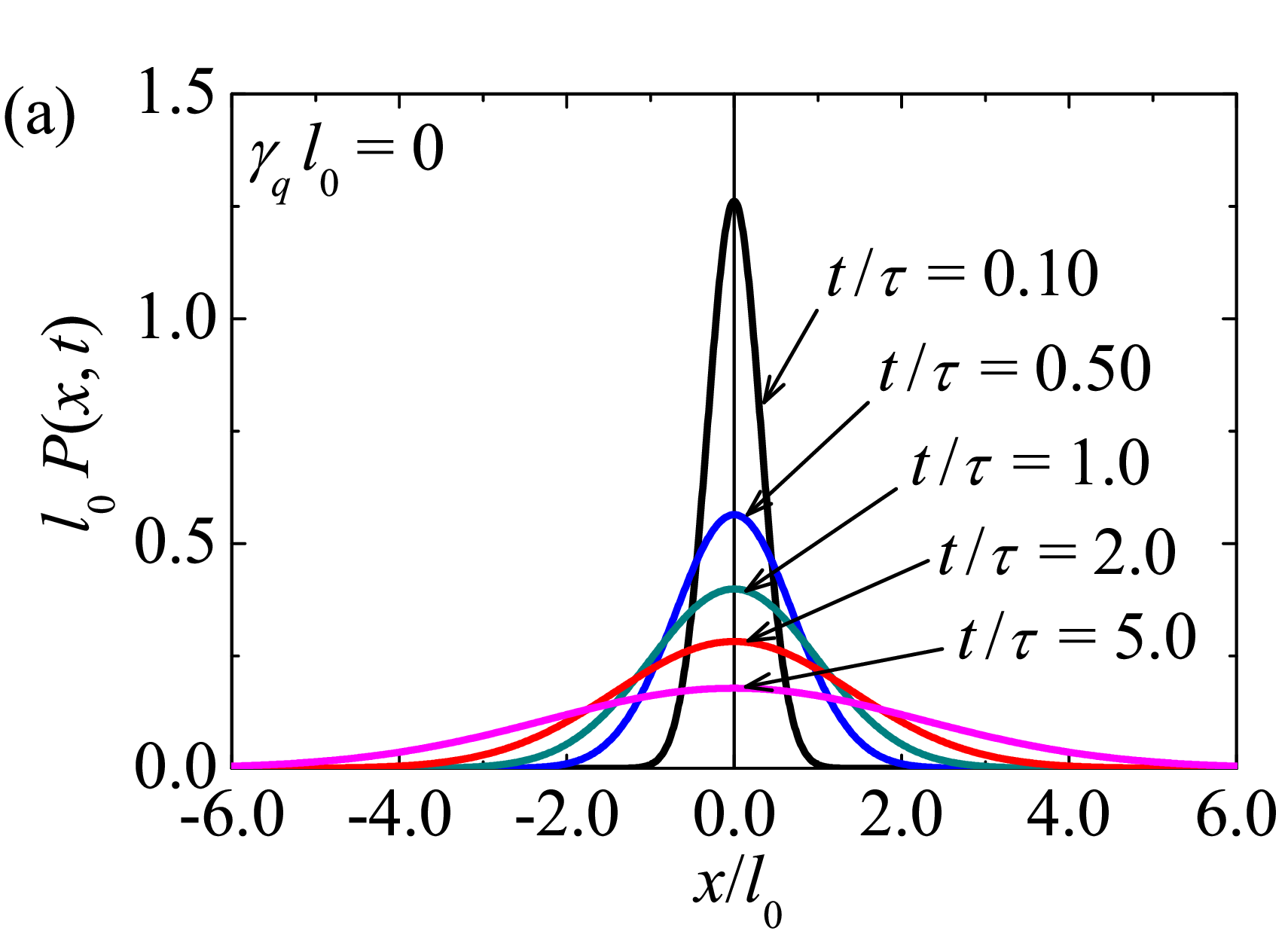}
\end{minipage}
\begin{minipage}[h]{0.32\linewidth}
\includegraphics[width=\linewidth ]{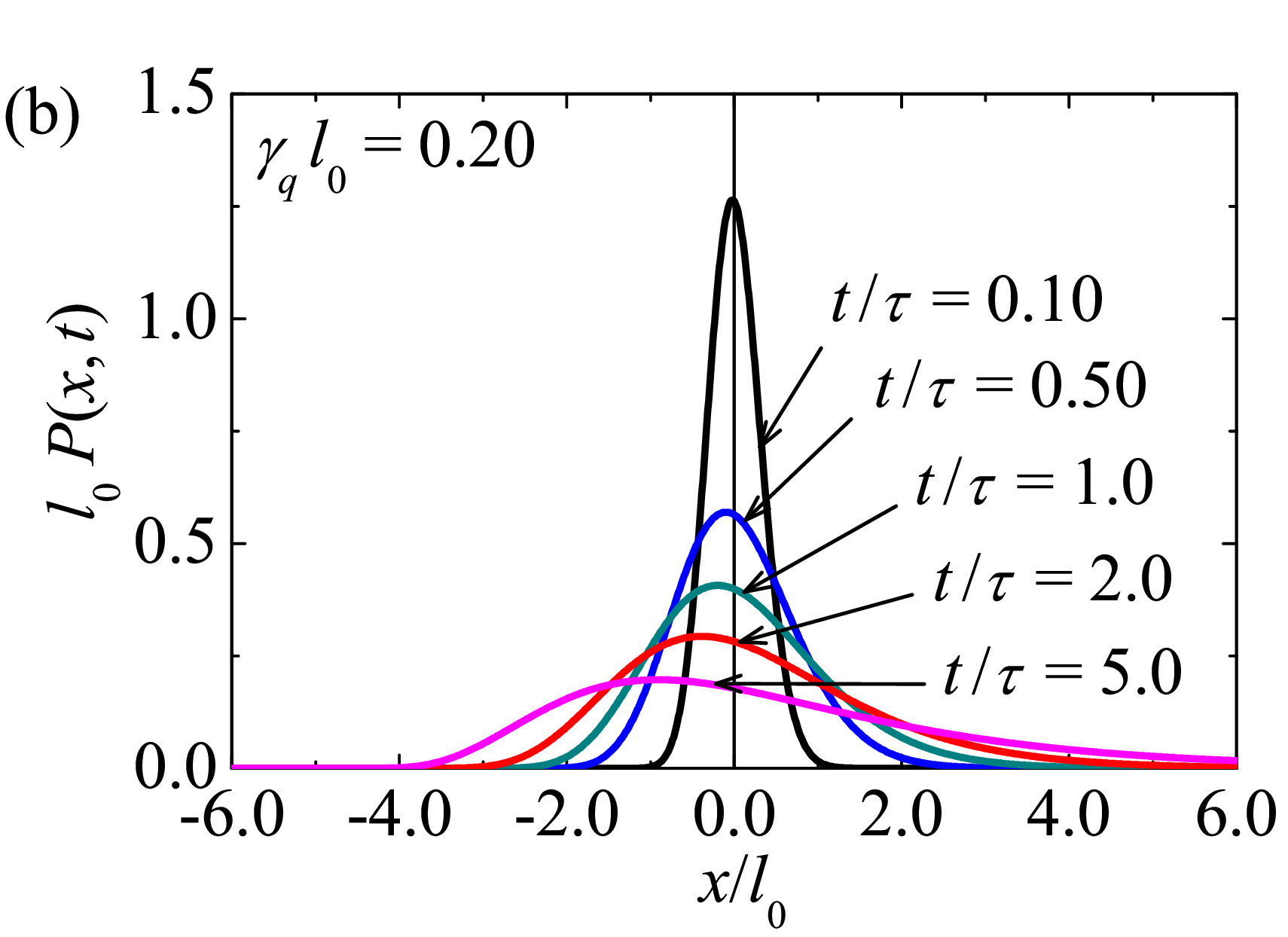}
\end{minipage}
\begin{minipage}[h]{0.32\linewidth}
\includegraphics[width=\linewidth]{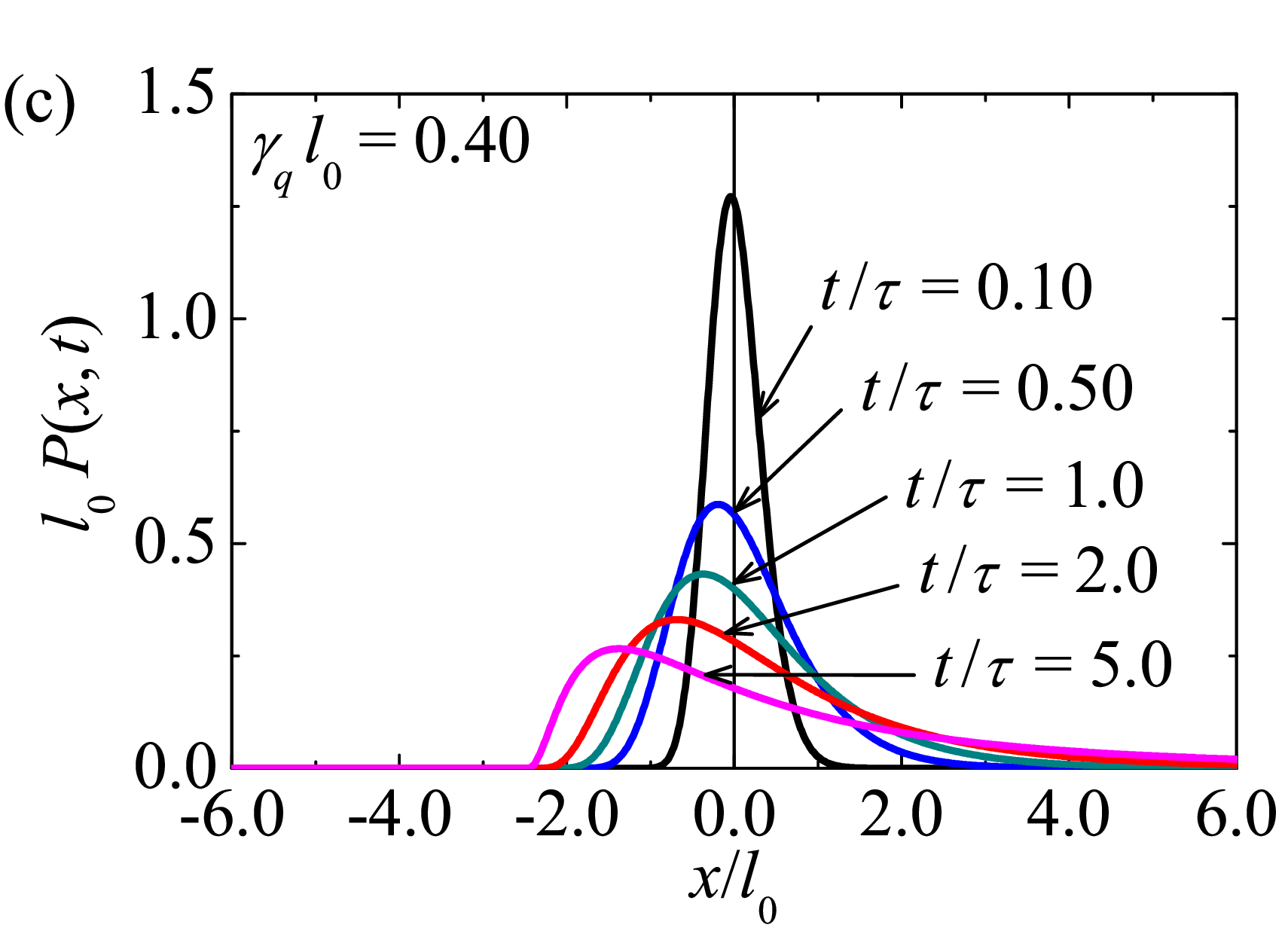}
\end{minipage}\\
\begin{minipage}[h]{0.32\linewidth}
\includegraphics[width=\linewidth]{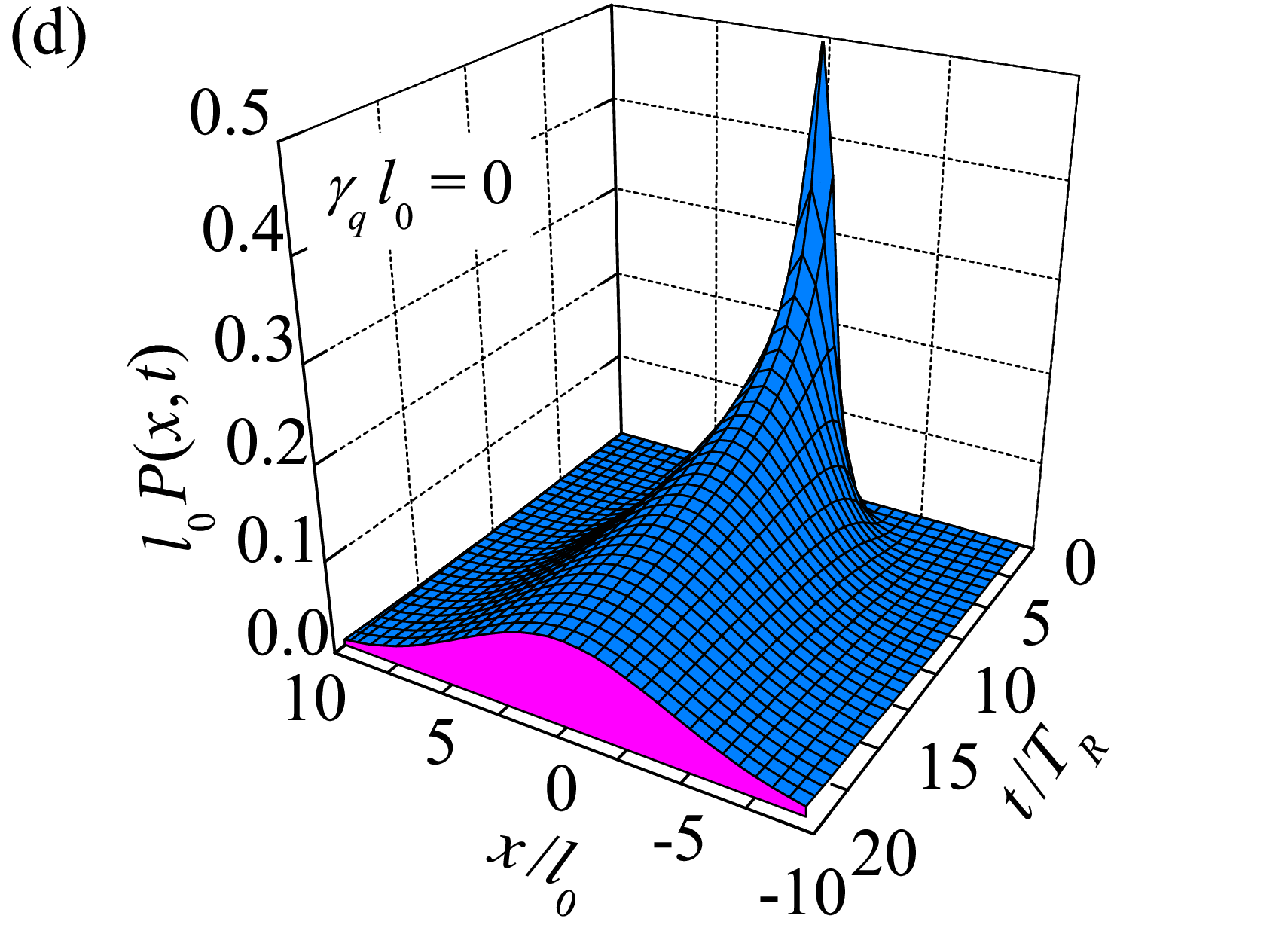}
\end{minipage}
\begin{minipage}[h]{0.32\linewidth}
\includegraphics[width=\linewidth ]{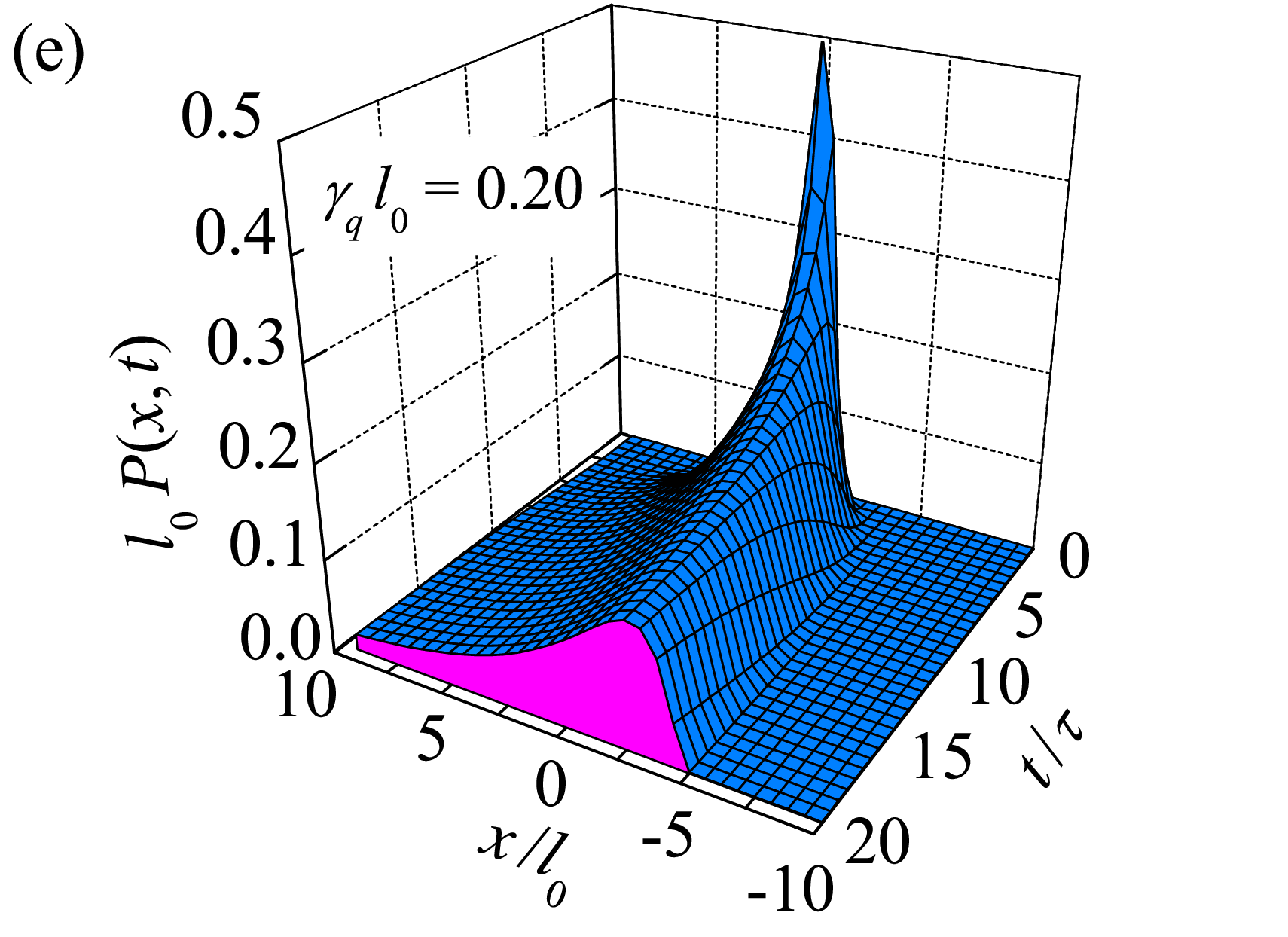}
\end{minipage}
\begin{minipage}[h]{0.32\linewidth}
\includegraphics[width=\linewidth]{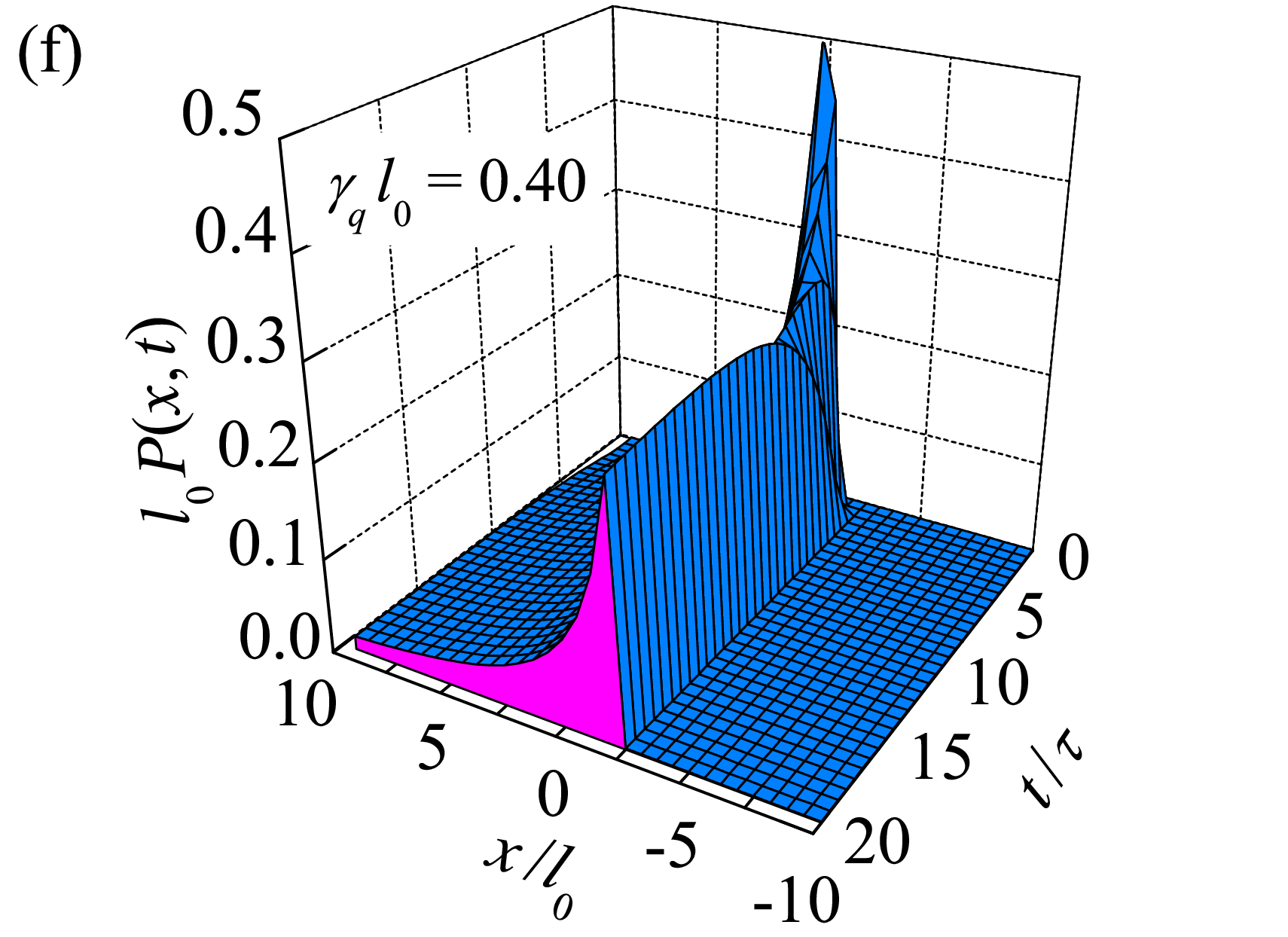}
\end{minipage}\\
\caption{\label{fig:figure1}
2D (upper line) and 3D (bottom line) representations of the solutions
$P(x, t)$ of the inhomogeneous FPE for free particle with parameters
$\gamma_q l_0 = 0$ (usual case), $0.2$ and $0.4$.
An asymmetry with respect to $x=0$ increases with $\gamma_q l_0$.
The divergence of Eq.~(\ref{eq:m(x)}) for
$x \rightarrow x_\textrm{d}=-1/\gamma_q$ implies a diffusion
limited to the interval $x < x_\textrm{d}$.
Notice that the space and time axis of the 3D plots are inverted
(increase from right to left), for a better visualization.
}
\end{figure*}

\begin{figure}[htb]
\begin{minipage}[h]{0.70\linewidth}
\includegraphics[width=\linewidth]{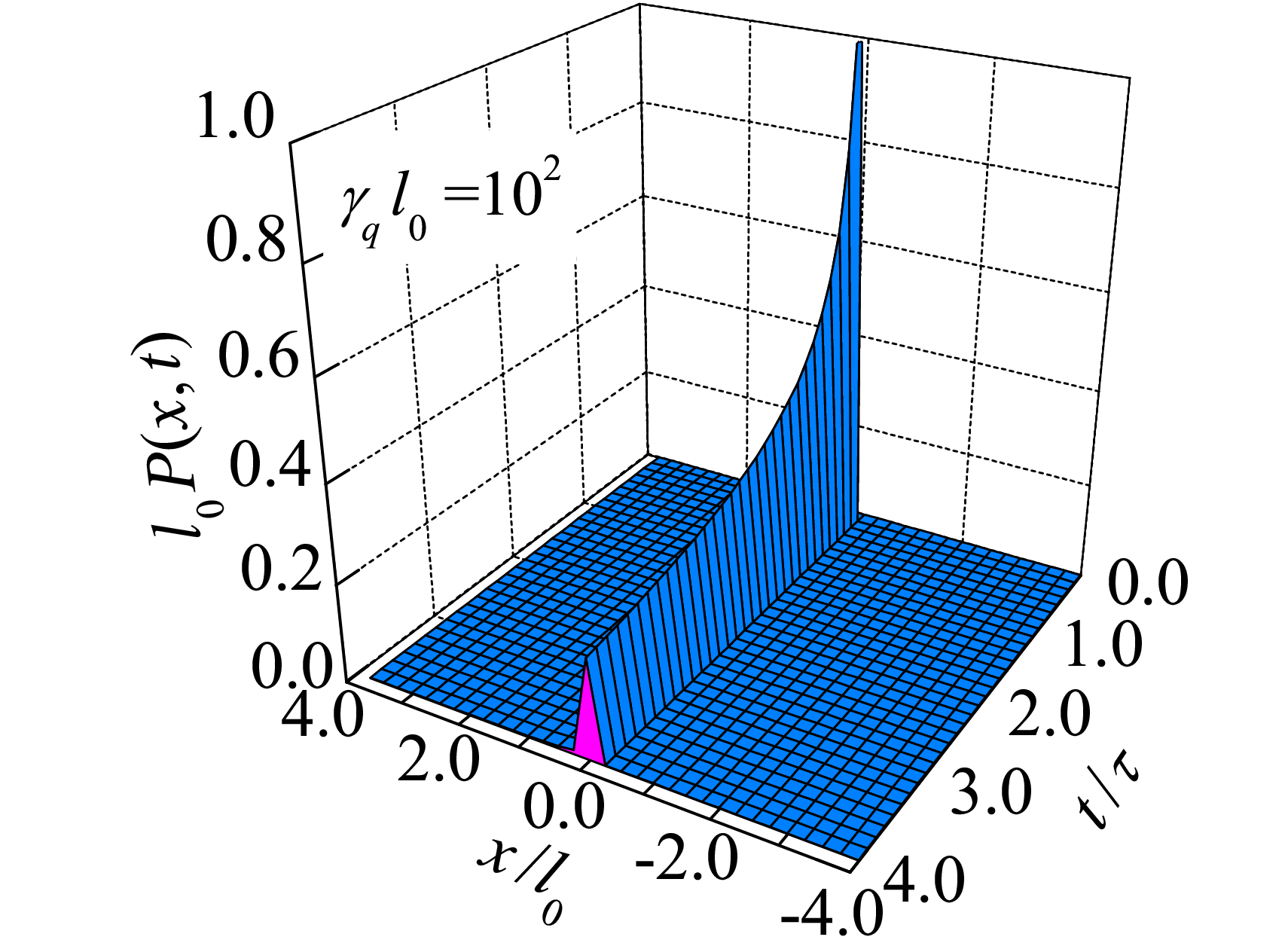}
\end{minipage}
\caption{\label{fig:figure2}
3D representation of $P(x, t)$ for $\gamma_q l_0=10^2$
showing that diffusion is stopped at $x=0$
for sufficiently
high values of $\gamma_q l_0$
(illustrated with
$x_{\textrm{d}}=-10^{-2})$ for a better visualization.
}
\end{figure}

Consider the $q$-deformed FPE for an infinite square well potential, where
$A(x) = 0$ for $ |x| \leq L/2$ and  $A(x) = \infty$ otherwise.
Using (\ref{eq:P_q(x)-stationary}) we obtain
$\mathcal{P}_q^{(\textrm{st})}(x) = C_q$ for
the stationary solution and from
the normalization  $1/C_q = \int_{-L/2}^{L/2} d_q x$ we have
\begin{equation}\label{eq:stationary-well}
 \mathcal{P}_{q}^{(\text{st})}(x)
 = \frac{\gamma_q}{\ln \left( \frac{1+\gamma_q L/2}{1-\gamma_q L/2}\right)}
 = \frac{1}{L_q},
\end{equation}
where $L_q$ is a deformed characteristic length.
The eigenfunctions of the associated FPE operator satisfy
\begin{equation}
 \label{eq:differential-equation-phi(x)}
 \mathcal{D}_q^2 \phi (x) = -k^2 \phi (x),
\end{equation}
with $k^2 = -2\Lambda/\Gamma$.
The confinement of the particle imposes $J(\pm\frac{L}{2})=0$, so
$\mathcal{D}_q\phi(\pm L/2) = 0$, and
thus, the solution of (\ref{eq:differential-equation-phi(x)})
is
\begin{equation}
 \phi_{q,n}(x) = \frac{1}{L_q}
             \cos \left[
                        \frac{k_{q,n}}{\gamma_q}
                         \ln \left(
                               \frac{1+\gamma_q x}{1 + \frac{1}{2}\gamma_q L}
                             \right)
                  \right],
\end{equation}
where $k_{q,n} = n\pi/L_q$, $n$ is a positive integer,
and the constant
$1/L_q$ has been chosen such that
$\phi_0(x) = \mathcal{P}_q^{(\text{st})}(x)$.
The general solution for $t=0$ is
\begin{equation}
 \mathcal{P}_q(x,0) =
  \sum_n c_n
         \cos \left[ \frac{k_{q,n}}{\gamma_q}
                     \ln \left(
                                \frac{1+\gamma_q x}
                                     {1 + \frac{1}{2}\gamma_q L}
                         \right)
              \right].
\end{equation}
The coefficients of the above expansion are obtained
from the following $q$-integrals
\begin{eqnarray}
 c_0 &=& \int_{-L/2}^{L/2} \mathcal{P}_q(x,0) d_q x, \\
 c_n &=& 2 \int_{-L/2}^{L/2} \mathcal{P}_q(x, 0) \phi_n(x) d_q x,
 \qquad (n \neq 0).
\end{eqnarray}
As usual, assuming a delta function for the initial condition
$P(x,0) = \mathcal{P}_q(x,0)/(1+\gamma_q x) = \delta(x)$,
we obtain $c_0 = 1/L_q $ and
$c_n = 2/L_q \cos[ k_{q,n}\gamma_q^{-1}\ln(1+\gamma_q L/2)]$
for $n \neq 0$. Thus, we have
\begin{eqnarray}
\label{eq:well-general-solution}
\mathcal{P}_q(x, t)  &=&
	\frac{1}{L_q}
	\Biggl\{ 1 + 2\sum_{n=1}^{\infty} \left(
	\cos \left[\frac{k_{q,n}}{\gamma_q}
			 \ln \left( 1+\frac{\gamma_q L}{2} \right) \right] \times
 	\right.
	\nonumber \\
	&&
	\left.
	\cos \left[\frac{k_{q,n}}{\gamma_q}
			 \ln \left( \frac{1+\gamma_q x}{1 + \frac{1}{2}\gamma_q L} \right) \right]
	e^{-t \Gamma k_{q,n}^2/2}  \right) \Biggr\},
\end{eqnarray}
with $x>-1/\gamma_q$ (here we are assuming $1+\gamma_qL/2>0$).
Consistently, when $\gamma_q \rightarrow 0$ the standard case is recovered:
\begin{eqnarray}
P(x, t)= \frac{1}{L}\left[
		1 + 2 \sum_{n=1}^{\infty} e^{-t \Gamma k_n^2/2} \cos(k_n x)
		\right],
\end{eqnarray}
with  $k_n = k_{1,n} = n\pi/L$.
In the limit of a large well, $L \rightarrow \infty$,
the general solution
(\ref{eq:well-general-solution}) takes the form
\begin{eqnarray}
\mathcal{P}_q(x,t) &=& \frac{2}{\pi} \lim_{L\rightarrow \infty}
						\int_{0}^{\infty}
		\left\{ \cos \left[\frac{k}{\gamma_q}
			 \ln \left( 1+\frac{\gamma_q L}{2} \right) \right] \times
		\right.
		\nonumber \\
		&& \qquad \qquad
		\left. \cos \left[\frac{k}{\gamma_q}
			 \ln \left( \frac{1+\gamma_q x}{1 + \frac{1}{2}\gamma_q L} \right) \right]
				 e^{-\Gamma k^2 t/2} \right\} dk
		\nonumber \\
		&=& \frac{1}{\sqrt{2\pi \Gamma t}}
			\Biggl\{
			\exp \left[ -\frac{\ln^2 (1+\gamma_q x)}{(2\Gamma t)\gamma_q^2}  \right]
			\nonumber \\
		&&
		+ \lim_{L\rightarrow \infty} \exp \left[ -\frac{1}{(2\Gamma t)\gamma_q^2}
			\ln^2\left(\frac{1+\gamma_q x}{1+\frac{\gamma_q L}{2}}\right)  \right] \Biggr\}.
\end{eqnarray}
The second term vanishes, then
\begin{equation}
 \label{eq:PDF-free}
 \mathcal{P}_q(x,t) =
 \frac{1}{\sqrt{2\pi \Gamma t}}
 \exp \left[ -\frac{\ln^2 (1+\gamma_q x)}{(2\Gamma t)\gamma_q^2}  \right].
\end{equation}
Recalling the deformed space
\begin{equation}
\label{eq:x_q(x)}
x_q (x) = \frac{\ln( 1+\gamma_{q} x)}{\gamma_q} ,
\end{equation}
and $\sigma^2(t)= \Gamma t$, Eq.~(\ref{eq:PDF-free})
can be recast as
\begin{equation}
 \label{estacionaria-qdeformed}
 \mathcal{P}_q(x,t) =
                      \frac{1}{\sqrt{2\pi \sigma^2(t)}}
                      \exp \left[
                                 -\frac{x_q^2(x)}{2\sigma^2(t)}
                           \right],
\end{equation}
that corresponds to a $q$-deformed solution of the free particle case.
The standard stationary solution (\ref{eq:estacionariasol-fokkerplanck})
is recovered at $q \to 1$.
Figures \ref{fig:figure1} and \ref{fig:figure2} illustrate
the solution (\ref{eq:PDF-free}) for some representative
values of the dimensionless parameter $\gamma_q l_0$.
As a consequence of the particular form of $m(x)$ (\ref{eq:m(x)}),
the diffusion is asymmetrical and the PDF is concentrated in a zone
near to the mass asymptote
$x_\textrm{d} = -1/\gamma_q$,
where the particle tends to have an infinite mass.
By contrast, in the region $x\geq -x_\textrm{d}$ the PDF
rapidly tends to zero as time evolves.
Moreover, as $\gamma_q l_0$ increase, the particle becomes more localized
at $x=0$ because the region where the PDF
can diffuse becomes small, as shown in Fig.~\ref{fig:figure2}.
%

\begin{figure}[htb]
\centering
\begin{minipage}[h]{0.70\linewidth}
\includegraphics[width=\linewidth]{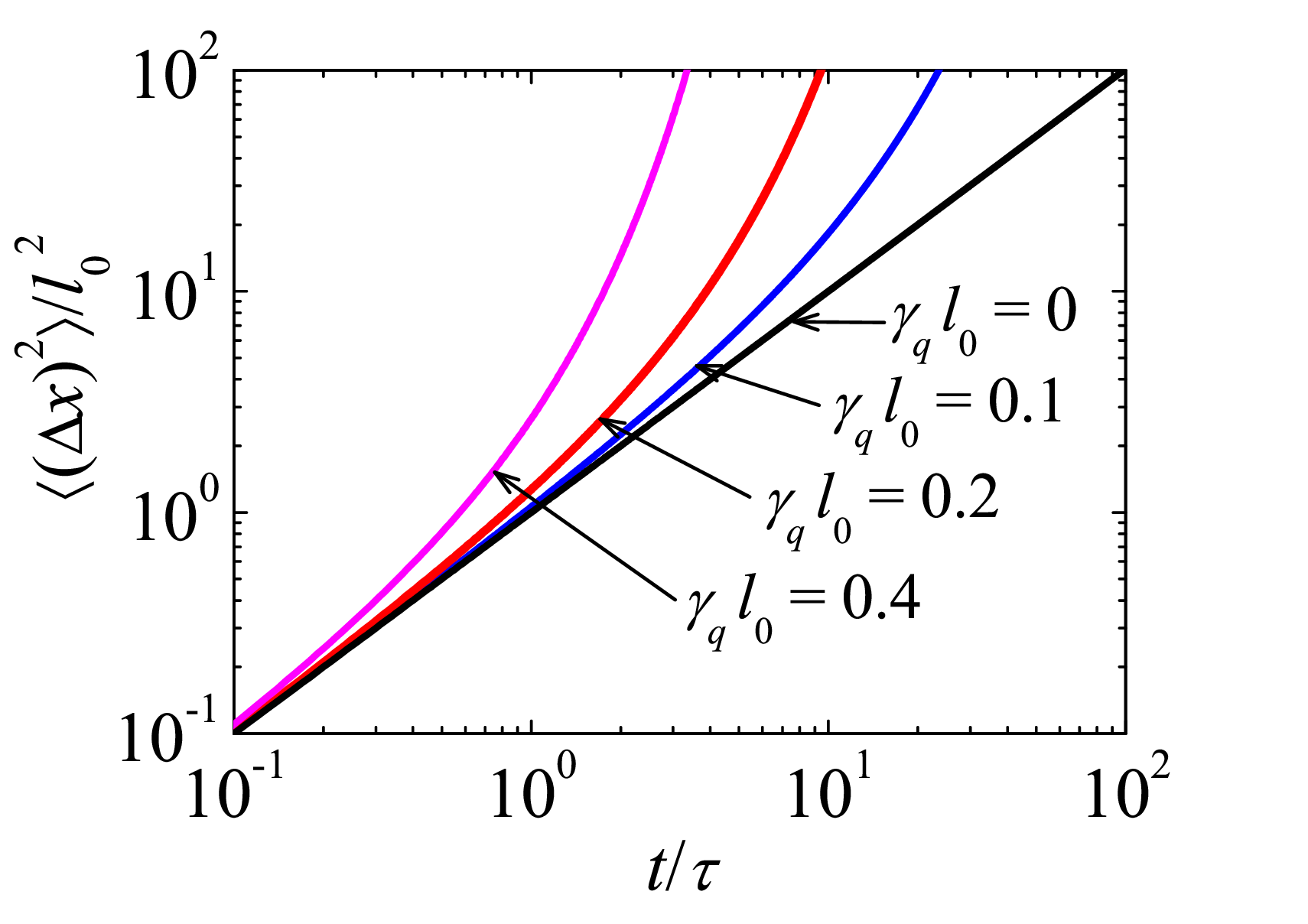}
\end{minipage}
\caption{\label{fig:figure3}
 Plot of
 $
  \langle (\Delta x)^2(t) \rangle
  = \langle x^2 (t) \rangle - \langle x(t) \rangle^2
 $
as a function of time for a free particle.
Normal diffusion behavior,
$\langle (\Delta x)^2 \rangle \approx \Gamma t$,
is observed for $t/\tau \ll 1$, $\forall \gamma_q l_0$,
and exponential hyperdiffusion,
$\langle (\Delta x)^2 \rangle \propto e^{2t/\tau}$, for $t/\tau \gg 1$.
}
\end{figure}

The transformation $x \rightarrow x_q$ in equation (\ref{eq:PDF-free})
leads the $n$-th moment of the distribution
\begin{equation}
 \langle x^n(t) \rangle
         = \int_{-\infty}^{\infty} x^n P(x, t) dx
         = \int_{-\infty}^{\infty} x^n \mathcal{P}_q(x, t) d_q x			
\end{equation}
into
\begin{eqnarray}
 \langle x^n(t) \rangle
         &=& \int_{-\infty}^{\infty}
                  \frac{1}{\sqrt{2 \pi \Gamma t}}
                  \frac{x^n}{1+\gamma_q x}
                  \exp \left[
                        -\frac{\ln^2 (1+\gamma_q x)}{(2\Gamma t)\gamma_q^2}
                       \right]
                  dx
         \nonumber \\
         &=& \int_{-\infty}^{\infty}
                  \frac{1}{\sqrt{2\pi \Gamma t}}
                  \left(
                         \frac{e^{\gamma_q x_q} -1}{\gamma_q}
                  \right)^n
                  e^{-{x}_q^2/(2\Gamma t)}
                  dx_q.
 \nonumber
\end{eqnarray}
The first and second moments are
\begin{subequations}
 \label{eq:x-moments}
 \begin{eqnarray}
  \langle x(t) \rangle &=&
          \frac{e^{(\Gamma t)\gamma_{q}^2/2}-1}{\gamma_q},
  \\
  \langle x^2(t) \rangle &=&
          \frac{e^{2(\Gamma t)\gamma_q^2}-2e^{(\Gamma t)\gamma_{q}^2/2}+1}
               {\gamma_q^2}.
 \end{eqnarray}
\end{subequations}
Figure \ref{fig:figure3} shows $\langle (\Delta x)^2(t) \rangle$
as a function of time for some values of $\gamma_q l_0$.
The spreading is hyperdiffusive,
i.e., faster than the superballistic power-law diffusion,
and exponentially increases for $t/\tau \gg 1$,
with a characteristic time $\tau =1/(\gamma_q^2\Gamma)$.
The normal diffusion is recovered for $\gamma_q\rightarrow0$,
corresponding to an infinite characteristic time $\tau$.

\subsection{\label{subsec:linear-potential}
            Confining potential with linear drift coefficient}

The $q$-deformed FPE  for $A(x) = -\alpha x$ is
\begin{equation}
 \label{eq:q-FPE-linear-potential}
 \frac{\partial \mathcal{P}_q (x, t)}{\partial t}
  = \alpha \mathcal{D}_{q} [x \mathcal{P}_q(x, t)]
    + \frac{\Gamma}{2} \mathcal{D}_{q}^2 {\mathcal{P}_q(x, t)}.
\end{equation}
In this case the associated effective potential
(\ref{eq:effective-potential}) is given by
\begin{equation}
 V_{\textrm{ef}}(x) =
 \frac{\alpha^2}{2\Gamma}x^2 - \frac{\alpha \gamma_q}{2}x - \frac{\alpha}{2}.
\end{equation}
The eigenfunctions $\psi_q(x)$ for the operator $\hat{\mathcal{K}}_q$
(see equation (\ref{eq:K_q-psi_q})) can be obtained from a comparison
with the solutions of the $q$-deformed time-independent Schr\"odinger
equation for a harmonic oscillator with frequency $\omega_0$
(for the usual case $q=1$) and electric charge $e$ in a uniform electric field
$\vec{\mathcal{E}} = \mathcal{E} \hat{x}$
\cite{Nascimento-Ferreira-Aguiar-Guedes-CostaFilho-2018}:
\begin{equation}
 \label{eq:q-osc-in-elec-field}
 - \frac{\hbar^2}{2m_0}\mathcal{D}_{q}^2\psi_{q}
 + \left(\frac{1}{2}m_0 \omega_0^2 x^2 - e\mathcal{E}x + V_0 \right) \psi_{q}(x)
 = E \psi_{q}(x),
\end{equation}
where $V_0$ is a constant.
The solutions of equation (\ref{eq:q-osc-in-elec-field})
in absence of an electric field has been studied
\cite{CostaFilho-Alencar-Skagerstam-AndradeJr-2013, Costa-Borges-2018},
the eigenfunctions and energies are obtained by means of a canonical point
transformation that maps the system into a Morse oscillator.
A similar transformation can be used for $\vec{\mathcal{E}} \neq 0$.
From the change of variables $\chi(s) = \psi_q(x(s))$
with $s(x)=\gamma_q^{-1}\ln[(1+\gamma_qx)/(1+\gamma_q x_0)]$
and $x_0 = e\mathcal{E}/(m_0\omega_0^2)$, it follows
\begin{equation}
 -\frac{\hbar^2}{2m_0} \frac{d^2{\chi}(s)}{ds^2}
 + \frac{m_0 \Omega_q^2}{2\gamma_q^2}(e^{\gamma_q s} -1)^2 \chi(s)
 = \widetilde{E} \chi (s).
\end{equation}
This equation corresponds to a quantum Morse oscillator
with frequency of small oscillations
$\Omega_q = \omega_0 (1+\gamma_q x_0)$
around the equilibrium position
and energy $\widetilde{E} = E -V_0 + e^2 \mathcal{E}^2/(2m_0\omega_0^2)$.
Consequently, the eigenfunctions of equation (\ref{eq:q-osc-in-elec-field}) are
\begin{equation}
 \label{eq:egeinfucntions-osc-elec-field}
 \psi_{q,n} (x) = \chi_n(s(x))
                 = A_n e^{-z(x)/2} [z(x)]^{\nu/2}
                   L_n^{(\nu)} (z(x)),
\end{equation}
where $z(x) = 2d(1+\gamma_q x)$, $d= m_0 \omega_0/(\hbar \gamma_q^2)$,
$\nu = 2d(1+\gamma_q x_0) - 1 - 2n > 0$, $A_n^2 = \nu \gamma_q n!/(\nu+n)!$,
and $L_n^{(\nu)}(z)$ are the associated Laguerre polynomials.
The energy eigenvalues of the equation (\ref{eq:q-osc-in-elec-field}) are
\begin{eqnarray}
 \label{eq:egeinvalues-osc-elec-field}
 E_n &=& V_0 -\frac{e^2\mathcal{E}^2}{2m_0\omega_0^2}
 + \hbar \omega_0 \left( 1+\frac{\gamma_q e\mathcal{E}}{m_0\omega_0^2}\right)
 \left( n+\frac{1}{2} \right)
 \nonumber \\
 && - \frac{\hbar^2 \gamma_q^2}{2m_0} \left( n+\frac{1}{2} \right)^2.
\end{eqnarray}
The number of bound states of the deformed oscillator is
$N_b = \lfloor d(1 + \gamma_q x_0) -1/2 \rfloor$, $\lfloor u \rfloor$
denoting the floor function, which tends to increase (decrease)
for $\gamma_q x_0 > 0$ ($\gamma_q x_0 < 0$)
in the presence of an external electric field.
The relations
$\hbar^2/m_0 = \Gamma$, $m_0 \omega_0^2 = \alpha^2/\Gamma$,
$e\mathcal{E} = \alpha \gamma_q/2$, and $V_0 = -\alpha/2$, lead to
\begin{eqnarray}
 \label{eq:psi-oscillator}
  \psi_{q,n}(x) &=& A_n e^{-\eta (1+\gamma_q x)}
  \left[ 2\eta (1 + \gamma_q x) \right]^{\eta - n}
  \nonumber \\ & & \times
  L_n^{(2\eta -2n)}(2\eta (1 + \gamma_q x)),
\end{eqnarray}
where $\eta =\alpha/(\Gamma \gamma_q^2)$
and $A_n^2 = 2(\eta-n)\gamma_q n!/(2\eta - n)!$.
The eigenvalues of $\hat{\mathcal{K}}_q$ are
\begin{eqnarray}
 \Lambda_n &=& -E_n
  = -\alpha n\left( 1 - \frac{\Gamma \gamma_q^2}{2\alpha} n \right),
\end{eqnarray}
with $\Lambda_n < 0$ for all $n\in\mathbb{N}$, except $\Lambda_0 = 0$.
The eigenfunctions of (\ref{eq:psi-oscillator}) are orthogonalized
through the deformed inner product
$\int_{-\infty}^{+\infty}\psi_{q,n}(x)\psi_{q,n}(x)d_q x =\delta_{n,m}$.
The coefficients $c_n$ of (\ref{eq:general-solution-psi})
with the initial condition
$P(x,0) = \mathcal{P}_q(x,0)/(1+\gamma_q x) = \delta(x)$
are $c_n=\psi_{q,n}(0)/\psi_{q,0}(0)$, so the general solution of
equation (\ref{eq:q-FPE-linear-potential}) results
\begin{equation}
 \label{eq:general-solution-linear-potential}
 \mathcal{P}_q(x, t) = \frac{\psi_{q,0}(x)}{\psi_{q,0}(0)}
                       \sum_{n} \psi_{q,n}(x)\psi_{q,n}(0) e^{-t\Lambda_n}.
\end{equation}
The summation in equation (\ref{eq:general-solution-linear-potential})
has the form of a quantum propagator for the Morse oscillator
\cite{Toutounji-2017}, from which we obtain its stationary solution
\begin{equation}
\label{eq:stationary-pdf}
 \mathcal{P}_q^{(\text{st})}(x) =
  \frac{
        \gamma_q
        \left[
              \frac{1}{\sigma_0^2 {\gamma}_q^2}
              (1+\gamma_q x)
              e^{-(1+\gamma_q x)}
        \right]^{ \frac{1}{\sigma_0^2 {\gamma}_q^2}}
       }
       {\left( \frac{1}{\sigma_0^2 {\gamma}_q^2} \right)!},
\end{equation}
with $\Gamma/(2\alpha) =\sigma_0^2$.
The transformations
$\gamma_q\rightarrow -\gamma_q$, $x \rightarrow -x$,
and $x_\textrm{d}\rightarrow-x_\textrm{d}$ are equivalent,
due to the asymmetry of the diffusion (see Fig.\ \ref{fig:figure1}),
which tends to concentrate the probability density around $x=x_\textrm{d}$.
Alternatively, the stationary solution can be obtained from
(\ref{eq:P_q(x)-stationary}) using $A(x)=-\alpha x$.

\begin{figure}[htb]
\centering
\begin{minipage}[h]{0.70\linewidth}
\includegraphics[width=\linewidth]{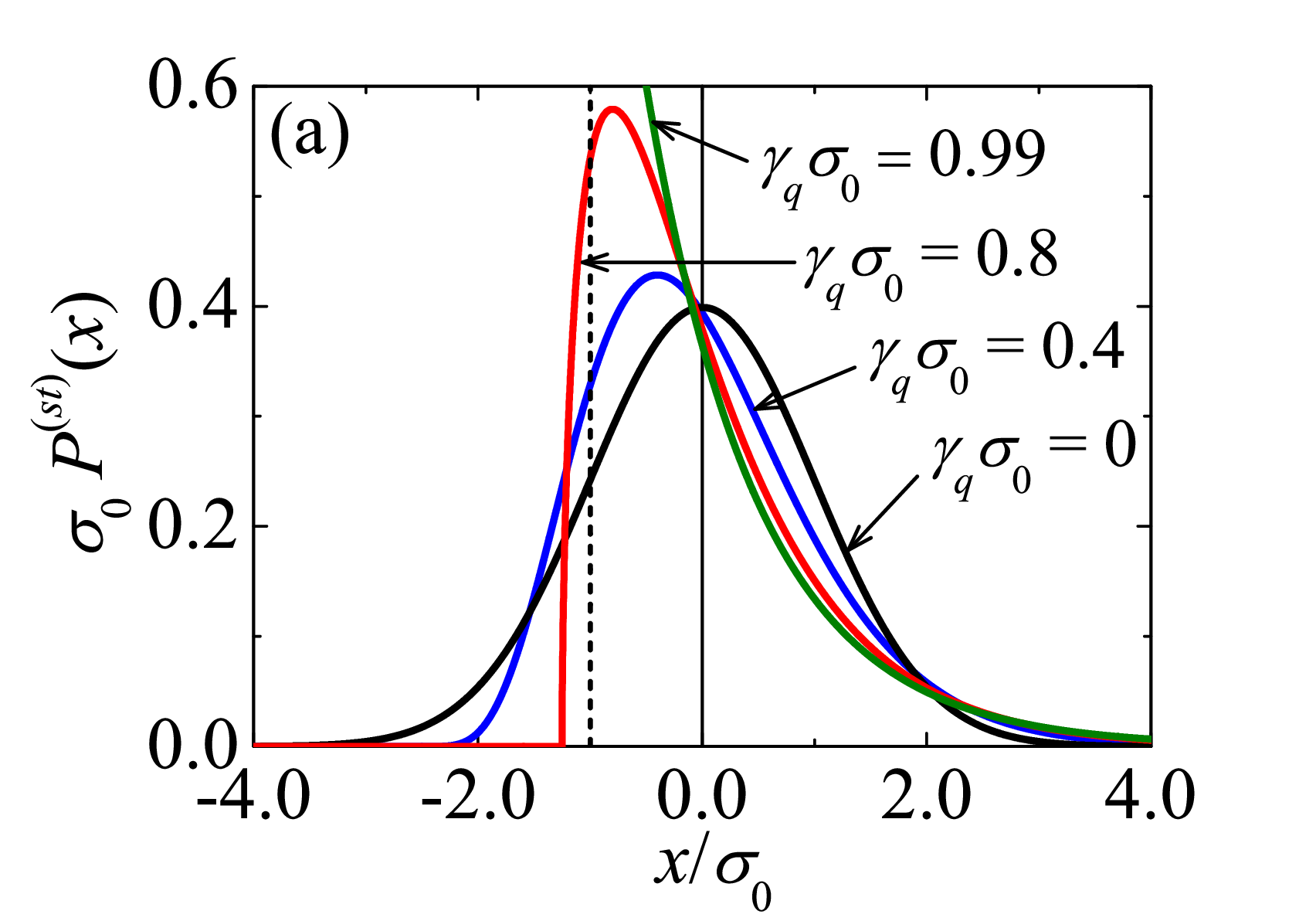}
\end{minipage}
\begin{minipage}[h]{0.70\linewidth}
\includegraphics[width=\linewidth]{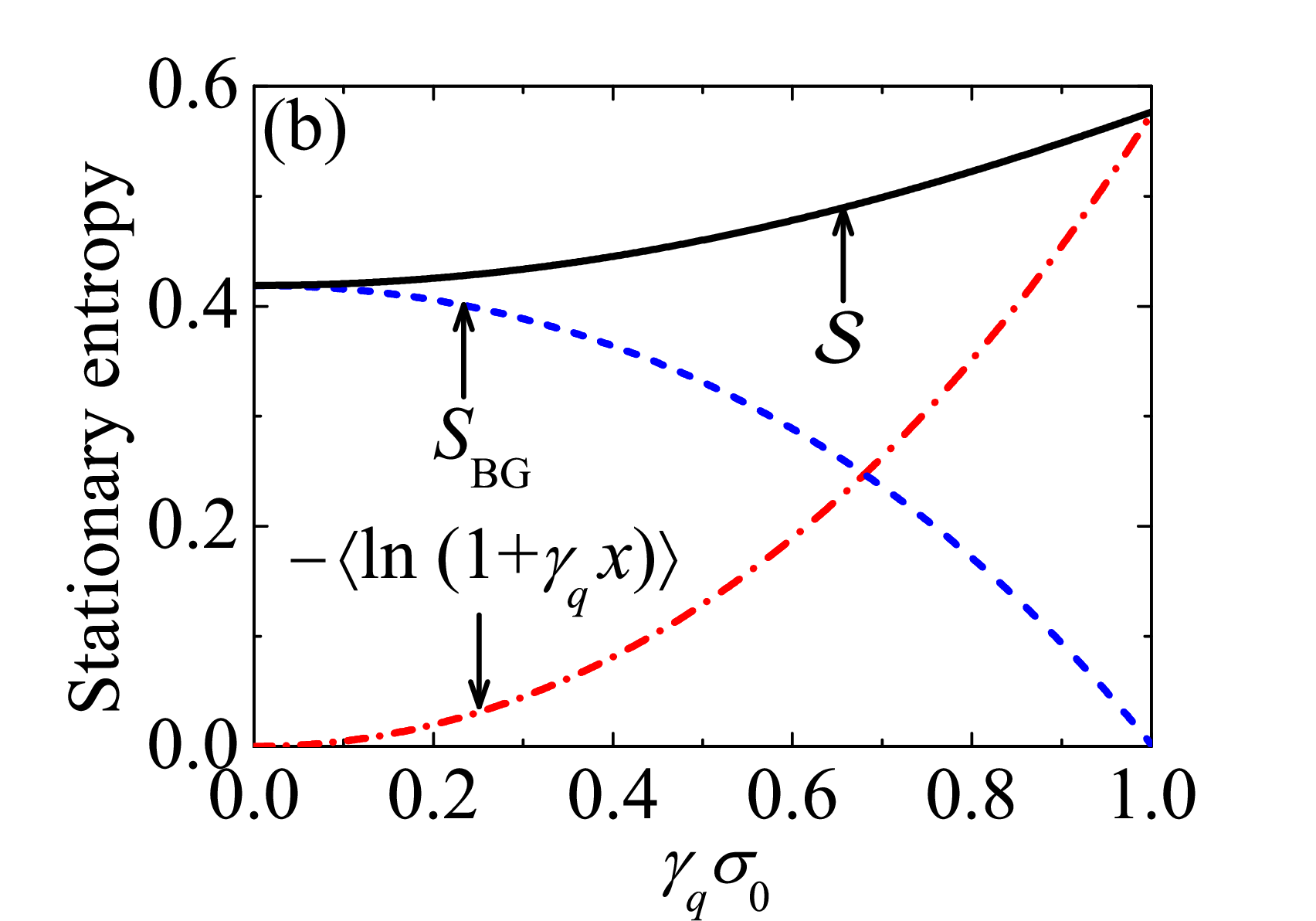}
\end{minipage}
\caption{\label{fig:figure4}
FPE for an inhomogeneous media with a linear potential.
(a) Stationary solution for different values of $\sigma_0\gamma_q$.
    Similarly to the free particle (see Fig.~\ref{fig:figure1}),
    asymmetry in the PDF is observed.
(b) Entropy $\mathcal{S}$ of the system (black line),
    the contribution of the entropy of the particles $S_{\text{BG}}$
    (blue dashed line),
    and the contribution of the medium, $-\langle \ln(1+\gamma_qx)\rangle$,
    (red dash-dot line),
    (see Eq.~(\ref{eq:deformed-entropy})),
    as a function of the inhomogeneity of the medium,
    controlled by $\gamma_q\sigma_0$.
}
\end{figure}

Figure \ref{fig:figure4}(a) shows some plots of the stationary solution
$P^{(\text{st})}(x) = \mathcal{P}_q^{(\text{st})} (x)/(1+\gamma_q x)$
for some values of $\sigma_0 \gamma_q$.
For $|\gamma_q| \rightarrow 1$ the PDF  (\ref{eq:stationary-pdf})
diverges at $x_\textrm{d} = -1/\gamma_q$.
Figure \ref{fig:figure4}(b) shows the deformed entropy
(\ref{eq:deformed-entropy})
as a function of $\gamma_q$ for the stationary PDF
along with the entropic contributions of the particles and the medium,
obtained by numerical integration.
Localization of particles at $x_\textrm{d}$ for $\gamma_q \sigma_0 \to 1$
implies $S_{\text{BG}} = - \int P \ln(\bar{\sigma} P) dx \to 0$
with $\bar{\sigma} = e = \text{constant}$.
The greater the value of the parameter $\gamma_q$,
the greater (smaller) is the entropic contribution
of the medium (particles) on the total entropy.

\section{\label{sec:discussion} Discussion and comparison with the literature}

Here follows a discussion of the formalism presented in the light of some
literature of inhomogeneous diffusion:
the van Kampen's approach
\cite{Landauer,vanKampen-87,vanKampen-book}
and the superstatistics
\cite{superstatistics,vanderStraeten}.
Also, we include two possible fluctuation theorems along with
an application of the deformed FPE to anomalous diffusion in optical lattices
\cite{Lutz,Beck-entropy}.

\subsection{Consistency with van Kampen's approach}

Our aim is to show that
the van Kampen's description of Sub-Section
\ref{subsec:van-Kampen-superstatistics}
can be expressed in terms of the deformed Fokker-Planck equation
(\ref{eq:fpe-inhomogeneous-deformed-derivative})
by means of a suitable choice
of the deformation $\kappa(x)$ for case in
which the functional form of the temperature $T$ and the
mobility of the particle $\mu$ are the same:
\begin{equation}
\label{eq:choice}
 \frac{T(x)}{T_0} = \frac{\mu(x)}{\mu_0}
                  = \frac{1}{\kappa(x)}
\end{equation}
with $T_0$ and $\mu_0$ their corresponding values
in the case of constant temperature and mobility.
By simple inspection between the equations (\ref{eq:P_kappa}),
(\ref{eq:fpe-inhomogeneous-deformed-derivative})
and (\ref{eq:van-Kampen-equation}), and the deformation
$\kappa(x)\equiv h(u)$ in \eqref{eq:generalized-linear-derivative}
$\mathcal{D}_{[\kappa]}=\frac{1}{\kappa(x)}\frac{d}{dx}$,
equation (\ref{eq:van-Kampen-equation}) can be rewritten as
\begin{equation}
 \label{eq:van-Kampen-deformed2}
 \frac{\partial \mathcal{P}_{[\kappa]}(x,t)}{\partial t}=
 \mathcal{D}_{[\kappa]}[\mu_0V'(x)\mathcal{P}_{[\kappa]}(x,t)]+
 \mu_0T_0\mathcal{D}_{[\kappa]}^2\mathcal{P}_{[\kappa]}(x,t)
\end{equation}
that is the deformed FPE (\ref{eq:fpe-inhomogeneous-deformed-derivative})
with the identification of the potential drift $A(x)$
and the constant $\Gamma$ as
\begin{eqnarray}
\label{eq:identification}
  A(x) &=& -\mu_0V'(x) \nonumber\\
  \Gamma/2 &=& \mu_0T_0.
\end{eqnarray}
We remark some consequences regarding
the connection between the van Kampen's diffusion equation
(\ref{eq:van-Kampen-equation}) and
the deformed FPE (\ref{eq:fpe-inhomogeneous-deformed-derivative}).
The first one is that the choice (\ref{eq:choice}) implies
\begin{equation}
 \label{eq:beta-pdm-relation}
 \beta(x)=\frac{1}{k_BT(x)}=\beta_0\sqrt{\frac{m(x)}{m_0}}.
\end{equation}
with
$\beta(x)=\beta_0$ corresponding to the constant temperature case,
thus linking the inverse of the temperature with the position-dependent mass.

Second remark, the entropic density
$s_{[\kappa]}(P)=-\frac{P}{\kappa}\ln\left(\frac{P}{\kappa}\right)$
satisfies $\frac{d^2s_{[\kappa]}}{dP^2}=-\frac{1}{\kappa P}<0$
and since the deformed stationary solution
$\mathcal{P}^{(\textrm{st})}(x)=\kappa(x)P^{(\textrm{st})}(x)$
maximizes $\mathcal{S}$
then $P^{(\textrm{st})}(x)$ (Section \ref{subsec:linear-potential})
also maximizes
\begin{equation}
 \label{eq:van-Kamppen-entropy}
 \mathcal{S}=S_{\textrm{BG}}-\langle \ln [T(x)/T_0] \rangle
\end{equation}
The expression (\ref{eq:van-Kamppen-entropy}) represents an entropy functional
for the existence of the deformed $H$-theorem
(Sub-Section \ref{subsec:deformed-theorem-H})
in an inhomogeneous medium with a position-dependent temperature $T(x)$.
The first term of Equation \eqref{eq:van-Kamppen-entropy}
has a microscopic nature (the probability density function),
while its second term depends on a macroscopic variable.
This would be considered as an inconsistency
within the usual statistical mechanics framework,
but within the superstatistics context, the second term is an average
over a continuous of canonical ensembles of temperatures $T(x)$.
The confining potential with linear drift of
Sub-Section \ref{subsec:linear-potential} exemplifies this point.
In the context of the van Kampen's equation (\ref{eq:van-Kampen-equation})
this case corresponds to an inhomogeneous media
with a linear temperature profile given by
equations (\ref{eq:q-deformation}) and (\ref{eq:choice})
and a external force $-V'(x)=-\frac{\alpha}{\mu_0}x$ ($A(x)=-\alpha x$)
with $\alpha,\mu_0\geq0$.
Figure \ref{fig:figure4}(b) shows the increase of the entropy $\mathcal{S}$
with the growth rate of the temperature
$\frac{1}{T_0}\frac{dT}{dx}=\gamma_q$, $S_{\textrm{BG}}$,
decreases, but the contribution of the medium sufficiently compensates,
and $\mathcal{S}$ increases with the inhomogeneity of the temperature.

Finally, if $T(x)/\mu(x)\neq\textrm{constant}$,
equation (\ref{eq:van-Kampen-equation}) can be expressed
by means of deformed derivatives. In fact, using
that
$\mathcal{P}_{[1/\mu]}(x,t)=P(x,t)\mu(x)/\mu_0$,
$\mathcal{P}_{[1/T]}(x,t)=P(x,t)T(x)/T_0$
and
$\mathcal{D}_{[1/\mu]}=\frac{\mu(x)}{\mu_0}\frac{d}{dx}$,
$\mathcal{D}_{[1/T]}=\frac{T(x)}{T_0}\frac{d}{dx}$,
the van Kampen's diffusion equation (\ref{eq:van-Kampen-equation})
is written as
\begin{eqnarray}
\label{eq:van-Kampen-general}
\frac{\partial \mathcal{P}_{[1/\mu]}(x,t)}{\partial t}&=&
\mathcal{D}_{[1/\mu]}[\mu_0V'(x)\mathcal{P}_{[1/\mu]}(x,t)]
\nonumber \\
&& + \mu_0 T_0 \mathcal{D}_{[1/\mu]}^2[\mathcal{P}_{[1/T]}(x,t)].
\end{eqnarray}
The general solution of this equation
is beyond the scope of the this work.

\subsection{Superstatistics and position-dependent mass Langevin equations}

A deep connection between the position-dependent Langevin equation
\eqref{eq:masslangevin-pdm}
and the superstatistics version \eqref{eq:Langevin-superstatistics}
can be given by considering $\lambda(x)=\lambda_0$
and multiply the left and right sides of \eqref{eq:masslangevin-pdm}
by $\dot{x}$, then we obtain
\begin{equation}\label{eq:connection-Langevin-deformada-1}
 \frac{d}{dt}\left(\frac{1}{2}m(x)\dot{x}^2\right)=-m(x)\lambda_0\dot{x}^2+(F(x)+R(t))\dot{x}.
\end{equation}
By means of the change of variable
\begin{equation}
\label{eq:change-variable}
x_{[\kappa]}(x)=\int^{x}\sqrt{\frac{m(x^{\prime})}{m_0}}dx^{\prime}
\end{equation}
the equation \eqref{eq:connection-Langevin-deformada-1} can be rewritten as
\begin{subequations}
\begin{align}
\label{eq:Langevin-superstatistics-PDM}
& \frac{d{v}_{[\kappa]}}{dt} = -\lambda_0 v_{[\kappa]}
					  +\frac{\overline{F}(x_{[\kappa]})}{m_0}
  +\sqrt{\frac{2\lambda_0}{m_0\beta (x_{[\kappa]})}}\overline{\xi}(t),
  \\
& \frac{d x_{[{\kappa}]}}{dt} = v_{[{\kappa}]}
\end{align}
\end{subequations}
with
\begin{subequations}
\begin{align}
\label{eq:identification-superstatistics}
& \overline{F}(x_{[\kappa]})
	= F(x(x_{[\kappa]}))\sqrt{\frac{m_0}{m(x(x_{[\kappa]}))}}
	= -\frac{dV(x(x_{[\kappa]}))}{dx_{[\kappa]}},
	\\
& \beta (x_{[\kappa]}) = \beta_0 \frac{m(x(x_{[\kappa]}))}{m_0}	
	  				   = \beta_0 \kappa^2(x(x_{[\kappa]})),
	\\
& \overline{\xi} (t)=\sqrt{\frac{m_0 \beta_0}{2\lambda_0}}R(t)
\end{align}
\end{subequations}
where $\beta_0$ denotes the standard case
$\beta(x_{[\kappa]})= \textrm{constant}$.
The set of equations \eqref{eq:Langevin-superstatistics-PDM}
is formally identical to \eqref{eq:Langevin-superstatistics},
thus giving a demonstration by first principles
of the superstatistics Langevin equation
in terms of a position-dependent mass particle.

The stationary solutions of the superstatistics
\eqref{eq:superstatistics-stationary}
and of the deformed FPE \eqref{eq:deformed-stationary-solutions}
along with the relationship
$\beta(x)=1/\kappa(x)$ from \eqref{eq:beta-pdm-relation}
indicate they are the same distribution.
Moreover, from the deformed stationary solution of the confining potential
\eqref{eq:stationary-pdf}, the equation \eqref{eq:P_kappa}
and by the same procedure for obtaining
the velocity distribution (\cite{vanderStraeten}, equation (16)) in the
overdamped limit, we obtain the distribution $f(\beta_q)$
for the $q$-deformation $\beta_q(x)=\frac{1}{1+\gamma_q x}$
\begin{equation}
\label{eq:inverse-Gamma}
f({\beta}_{q}) =
\frac{\beta_q^{-\alpha-1}}{\Gamma(\alpha)}\exp\left(-\frac{\theta}{\beta_q}\right),
\quad
\alpha = \theta = \frac{1}{\sigma_0^2\gamma_q^2},
\end{equation}
which is the inverse Gamma distribution of the example
$\beta(x)=\frac{1}{|x|+a}$ of \cite{vanderStraeten}
in the limit $a\rightarrow0$.
Also, from $\beta_q(x)$ other candidate for the force
$\frac{2}{\Gamma}A(x)$ can be obtained
by means of equation (18) of \cite{vanderStraeten}.

Position-dependent mass and superstatistical Langevin
equations (in $x$ and $y$ spaces respectively) \eqref{eq:masslangevin-pdm}
and \eqref{eq:Langevin-superstatistics-PDM},
are equivalent, maintaining the position and the velocity
at the same status level,
from which results the overdamped PDM Langevin equation \eqref{eq:masslangevin-pdm-2}
(or equivalently \eqref{eq:deformed-Langevin}),
in $x$ for $\lambda(x)\gg\tau^{-1}$.
Analogously, the van Kampen's FPE \eqref{eq:van-Kampen-equation}
is not equivalent to the superstatistical Langevin equation \eqref{eq:Langevin-superstatistics},
since the overdamped limit has not been taken in the latter.

\subsection{Work fluctuation theorems for position-dependent mass particle}

We outline two possible fluctuation theorems (FT)
\cite{Kubo-1966,FDT-Beck} in a position-dependent mass scenario
by reviewing some works on fluctuation theorems for
a dragged Brownian particle \cite{FT-Wang,FT-Cohen}.
For simplicity we restrict our discussion to the deformation
\eqref{eq:q-deformation}.
In order to apply the FT theorem \cite{FT-Cohen}
and inspired by the experiment of Wang,
we consider a one-dimensional Brownian particle of  constant mass $m_0$
in a medium of friction $\lambda_0$ and temperature $T_0$
in the deformed frame $x_q$ (\ref{eq:x_q(x)}),
and subjected to a force $F(x_q,{x}^{*}_q(t))=-k(x_q-{x}^{*}_q(t))$
with an arbitrary time-dependent position $x^{*}(t)$.
Let us denote $W_\tau$ and $\overline{W}_\tau$
the works done on the system during a time $\tau$, with $\tau$ the time scale
of the fluctuations in the spaces $x$ and $x_q$ respectively.
By means of the transformation (\ref{eq:x_q(x)})
it is immediate to show that the overdamped position-dependent mass
Langevin equation (\ref{eq:masslangevin-pdm-2}) is equivalent to
\begin{equation}
 \label{eq:FDT-1}
 \dot{x}_q
  = -\frac{x_q-{x}^{*}_q(t)}{\tau_r} + \xi(t)
\end{equation}
with $\tau_r=\lambda_0/k$ the relaxation time, $\langle \xi(t)\rangle=0$ and
$\langle \xi(t)\xi(t^{\prime})\rangle=2k_BT_0\lambda_0\delta(t-t^{\prime})$
and a force
$F(x,x^{*}(t))=(1/\gamma_q)\ln((1+\gamma_qx)/(1+\gamma_qx^{*}(t)))$.
Under these conditions, from \eqref{eq:FDT-1} the work FT
of the equation (31) of \cite{FT-Cohen} in the space $x_q$ follows
\begin{equation}
 \label{eq:FDT-2}
 \frac{P(\overline{W}_\tau)} {P(-\overline{W}_\tau)}
 = e^{\overline{W}_\tau},
\end{equation}
with $P(\overline{W}_\tau)$ the probability distribution of
$\overline{W}_\tau$, constructed by measuring
$\overline{W}_\tau$ over time intervals $\tau$ \cite{FT-Wang}.
By the definition of equation (5) of \cite{FT-Cohen},
\begin{eqnarray}
 \label{eq:work}
 \overline{W}_\tau &=& \beta_0 \int_{0}^{\tau} dt \;
                        {v}_q^{*}(t) \;
                        [-k(x_q(t)-x_q^{*}(t))]
 \nonumber\\
                    &=& \beta_0\int_{0}^{\tau} dt \;
                        \frac{v^{*}(t)}{1+\gamma_q x^{*}(t)} \;
                        F(x(t),x^{*}(t))
 \nonumber\\
                    &=& \langle W_\tau \rangle_{q},
\end{eqnarray}
with ${v}_q= v/(1+\gamma_q x)$.
The probability of the work done must be the same on both
spaces $x$ and $x_q$, so
$P(W_\tau)dW_\tau=P(\overline{W}_\tau)d\overline{W}_\tau$,
from which follows
$P(\overline{W}_\tau)/P(-\overline{W}_\tau)=P(W_\tau)/P(-W_\tau)$.
Then, from \eqref{eq:work} we recast the work FT
\eqref{eq:FDT-2} in standard space $x$,
\begin{equation}\label{eq:first-FDT}
 \frac{P(W_\tau)}{P(-W_\tau)} = e^{\langle W_\tau \rangle_{q}},
\end{equation}
that constitutes a first version of the work FT
\cite{FT-Cohen} with
$W_\tau$ averaged by the deformation (\ref{eq:q-deformation}).
We can provide a second (stationary state) version of the
work FT,
now by measuring the ratio $P(W_\tau)/P(-W_\tau)$ over single trajectories in
a stationary state of the type \eqref{eq:superstatistics-stationary},
that is, by dividing a stationary trajectory of total time $t$
in a sequence of $M\gg1$ time intervals with duration
$\tau$ ($t_{i+1}-t_i=\tau$) and initial times $t_1,\ldots,t_M$
\cite{FT-Cohen}.
This corresponds to the stationary state fluctuation theorem (SSFT)
\cite{FT-Cohen},
as a case of the FT in the long term regime $t \gg \tau$.
For the constant velocity case, $v^{*}(t)=v_0$
of $x^{*}(t)$
\cite{FT-Wang,FT-Cohen},
it follows the time scale $\tau_L=1/(\gamma_qv_0)$,
during which the deformation \eqref{eq:q-deformation} varies.
The time scales ordering $\tau<\tau_L\ll t$
implies the factor $1/(1+\gamma_qx^{*}(t))$
in \eqref{eq:work}
is approximately constant,
so from \eqref{eq:first-FDT} we obtain
\begin{subequations}
 \label{eq:second-FDT}
 \begin{align}
 & \frac{P(W_\tau)}{P(-W_\tau)}=e^{\beta(x^{*}(t_0)) W_\tau}
 \\
 & \beta(x^{*}(t_0))=\beta_0/(1+\gamma_qx^{*}(t_0))
 \end{align}
\end{subequations}
with $t_0\in(0,\tau)$,
that can be considered a manifestation of the superstatistics work FT
of \cite{FDT-Beck} linked with position-dependent mass systems,
where we have used the identification \eqref{eq:beta-pdm-relation}.
Employing \eqref{eq:second-FDT} we can derive an expression
for the expectation of the ratio $P(W_\tau)/P(-W_\tau)$
for the inverse Gamma
distribution \eqref{eq:inverse-Gamma}
of the confining potential case.
Over a stationary trajectory in the long term regime,
by averaging \eqref{eq:second-FDT} with
\eqref{eq:inverse-Gamma}
we obtain
\begin{equation}
 \label{eq:deformed-FDT-confining-potential}
  \left\langle \frac{P(-W_\tau)}{P(W_\tau)} \right\rangle
  = \frac{2}{\Gamma(\alpha)} (\alpha W_\tau)^{\alpha/2}
    K_{\alpha}(2\sqrt{\alpha W_\tau}),
\end{equation}
with
$\alpha=1/(\sigma_0^2\gamma_q^2)$
and
$K_\nu(x)$ is the modified Bessel function of the second kind.
The average of the probability ratio \eqref{eq:deformed-FDT-confining-potential}
asymptotically decays as a power law $W_\tau^{-\alpha/2}$
for small values of $W_\tau$,
while for large $W_\tau$,
it decays exponentially as if there were no $\beta$ fluctuations.
We could extrapolate the validity of the FT for a Morse potential force,
that is, by making the substitution
$-(1/\gamma_q)[\exp(\gamma_q(x_q-x_q^{*})) -1]$
$\rightarrow$ $(x_q - x_q^{*})$
in \eqref{eq:FDT-1}
with a relaxation time $\tau_r=\lambda_0\gamma_q/D$
and $D$ the dissociation constant of the Morse
potential\footnote{Not to be confused with the diffusion coefficient,
that appears in others parts of this paper.}.
In the long term regime, the confining potential decays as a power law
or as an exponential, for small or large $W_\tau$, respectively.
This behavior follows from \eqref{eq:deformed-FDT-confining-potential},
along the same steps that lead to \eqref{eq:second-FDT}.
A possible test for equation (\ref{eq:deformed-FDT-confining-potential})
is the experiment referred to in Ref.~\cite{FT-Wang}
in the long term regime with temperature and mobility profiles given by
equations (\ref{eq:q-deformation}) and (\ref{eq:choice}),
together with the condition $\tau_L=1/(\gamma_qv_0)>\tau$.
%

\subsection{Anomalous diffusion in optical lattices}

Other important case of inhomogeneous diffusion has been investigated in
optical lattices \cite{Lutz}, whose relevance against others counterparts lies
in the fact that its optical periodic potential is completely known,
thus allowing to control it in a precise way.
In this regard, an intermediate atomic transport regime can be identified,
between diffusive motion and ballistic motion,
in which anomalous diffusion occurs and the dynamics is adequately described
by nonextensive statistics \cite{Lutz,Beck-entropy}.
In this regime, the atom--laser interaction in the optical lattice
is governed by a quantum master equation
whose spatial averaging gives the Rayleigh equation for the Wigner function
$W(p,t)$,
\begin{equation}
\label{eq:Rayleigh}
 \frac{\partial W(p,t)}{\partial t}
  =
 - \frac{\partial}{\partial p}[K(p)W(p,t)]
 + \frac{\partial}{\partial p}
                  \left[D(p)\frac{\partial W(p,t)}{\partial p}\right],
\end{equation}
where the functions $K(p)$ and $D(p)$ are the drift
(cooling force, the Sisyphus effect)
and diffusion (stochastic momentum fluctuations of $p$) coefficients.
Our purpose is to show that Rayleigh equation (\ref{eq:Rayleigh})
can also be expressed as a particular deformed FPE
(\ref{eq:fpe-inhomogeneous-deformed-derivative})
for the stationary case $\frac{\partial W(p,t)}{\partial t}=0$.
Noticing that the diffusion coefficient $D(p)$
defines the deformed derivative
(see \eqref{eq:generalized-linear-derivative} with $h(u) \equiv 1/D(p)$)
$\mathcal{D}_{1/D}=\frac{D(p)}{D_0}\frac{\partial}{\partial p}$
(with $D_0$ corresponding to fluctuations of photon emissions \cite{Lutz})
and by making $W(p,t)=\overline{W}(x,p)D(p)$,
then $W(p,t)$ can be interpreted as a deformed version of $\overline{W}(x,p)$,
i.e., $W(p,t)=\mathcal{\overline{W}}_{[1/D]}(x,p)$
(according to (\ref{eq:P_kappa})).
Thus, we recast (\ref{eq:Rayleigh}) for the stationary case as
\begin{equation}
 \label{eq:stationary-deformed-Rayleigh}
  0 = -\mathcal{D}_{[1/D]}[K(p)\mathcal{\overline{W}}_{[1/D]}(p)]
      + D_0 \mathcal{D}_{[1/D]}^2 \mathcal{\overline{W}}_{[1/D]}(p),
\end{equation}
which is entirely expressed in the deformed space
$d_{[1/D]}p=\frac{D_0}{D(p)}dp$
with constant diffusion coefficient $D_0$.
Moreover,
it follows from (\ref{eq:stationary-solution-deformed-space})
its stationary solution,
\begin{eqnarray}
 \label{eq:Rayleigh-stationary-solutions}
 W^{(\textrm{st})}(p)
   &=& \mathcal{\overline{W}}_{[1/D]}^{(\textrm{st})}(p)
 \nonumber\\
   &=& C\exp\left(\frac{1}{D_0}\int^{p}K(p^{\prime})d_{[1/D]}p^{\prime}\right)
 \nonumber\\
   &=& C\left[1-\beta(1-q)p^2\right]^{1/(1-q)},
\end{eqnarray}
which is the Tsallis distribution (equation (5) of \cite{Lutz}),
with $K(p)/D(p)=\frac{2\beta p}{1-\beta(1-q)(p)^2}$.

\section{\label{sec:conclusions} Conclusions}

\squeezetable
\begin{table*}[htb]
\caption{
\label{tab:q-deformed-equations}
Linear and nonlinear deformed Fokker-Planck
and Schr\"odinger equations.
}
\centering
\begin{ruledtabular}
\begin{tabular}{l | p{0.2\linewidth} | p{0.35\linewidth} | p{0.35\linewidth}}
%
&
\centering
{deformed derivative}
&
\qquad \quad
deformed Fokker-Planck equation
&
\qquad \quad
deformed Schr\"odinger equation
\\ \hline & & \\
Linear
&
\centering
$
\displaystyle
\mathcal{D}_{q}f(u) = [1+(1-q)u]\frac{df}{du}
$
&
\centering
$
\displaystyle
\begin{array}{rcl}
\displaystyle
\frac{\partial \mathcal{P}_q(x, t)}{\partial t} &=&
	-\mathcal{D}_{q,x}[A(x) \mathcal{P}_q(x,t)] \\
 	&&
\displaystyle
	+\frac{\Gamma}{2}
	\mathcal{D}_{q,x}^2 \mathcal{P}_q(x,t)
\end{array}
$	
&
\quad \quad
$
\begin{array}{rcl}
\displaystyle
i\hbar \frac{\partial \Psi_q (x,t)}{\partial t}
		& = &
		\displaystyle
		-\frac{\hbar^2}{2m_0} \mathcal{D}_{q,x}^2 \Psi_q (x,t)		
        \\
		& & + V(x) \Psi_q (x,t)
\end{array}
$
\\
&
\qquad \qquad
(Eq.~(\ref{eq:borges-linear-derivative-q-diff}))
&
\qquad \qquad
(Eq.~(\ref{eq:q-FPE}), proposed in this work)
&
\qquad \qquad
(Eq.~(\ref{eq:SE-pdm-bis}),
 proposed in \cite{CostaFilho-Almeida-Farias-AndradeJr-2011})
\\ \hline & & \\
Nonlinear
&
\centering
$
\displaystyle
\widetilde{\mathfrak{D}}_{q}f(u) = [f(u)]^{1-q}\frac{df}{du}
$
&
\centering
$
\displaystyle
\begin{array}{rcl}
\displaystyle
\widetilde{{\mathfrak{D}}}_{q,t} P_q(x,t) &=&
	-\widetilde{{\mathfrak{D}}}_{q,x}[A(x) P_q(x,t)] \\
 	&&
\displaystyle
	+\frac{\Gamma}{2}
	\widetilde{{\mathfrak{D}}}_{q,x}^2 P_q(x,t)
\end{array}
$	
&
\quad \quad
$
\begin{array}{rcl}
\displaystyle
i\hbar \widetilde{{\mathfrak{D}}}_{q,t} \Phi_q (x,t)
& = &
\displaystyle -\frac{\hbar^2}{2m_0}\widetilde{{\mathfrak{D}}}_{q,x}^2 \Phi_q (x,t)		
\\
&   & + V(x) \Phi_q (x,t)
\end{array}
$
\\
&
\qquad \qquad
 (Eq.~(\ref{eq:nobre-nonlinear-derivative}))
&
\qquad \qquad
(Eq.~(\ref{eq:nonlinear-FPE-and-derivative}),
 proposed in \cite{Plastino-Plastino-1995})
&
\qquad \qquad
(Eq.~(\ref{eq:SE-nobre-bis}),
proposed in \cite{Nobre-RegoMonteiro-Tsallis-2011})
\\
\end{tabular}
\end{ruledtabular}
\end{table*}

Quantum and classical formalisms properly deformed
to account for systems with position-dependent effective mass
recently addressed in the literature
\cite{CostaFilho-Almeida-Farias-AndradeJr-2011,
      CostaFilho-Alencar-Skagerstam-AndradeJr-2013,
      Costa-Borges-2014,
      Costa-Borges-2018,
      Costa-Gomez-2018,
      Nascimento-Ferreira-Aguiar-Guedes-CostaFilho-2018}
have been studied
for which derivative operators are replaced by their deformed forms.
Table I displays the whole picture,
exhibiting deformed versions of Fokker-Planck and Schr\"odinger equations,
and the gap fulfilled by the present work.
The linearity and the nonlinearity of the equations are rephrased
by linear and nonlinear versions of deformed derivatives.
We summarize our contributions as follows.

(i)
Two deformed derivatives have been generalized into a unified framework
within an arbitrary deformation space $h(x)$,
Eqs.~(\ref{eq:generalized-linear-derivative})
and (\ref{eq:generalized-nonlinear-derivative}).
This scenario allows to obtain a linear deformed Fokker-Planck equation
that is equivalent to the corresponding FPE
in an inhomogeneous media with a position-dependent mass
along with dumping and diffusion coefficients
as a function of the employed deformation.

(ii)
The deformation carries pieces of information
about the inhomogeneity of the medium,
as a consequence of the equivalence between the FPE in an inhomogeneous medium
with position-dependent mass
and a deformed FPE in a homogeneous medium with constant mass.

(iii)
There is a connection between the molecular and
the macroscopic (diffusion) deformed descriptions,
given by the Langevin
(\ref{eq:deformed-Langevin}) and the Fokker-Planck
(\ref{eq:q-FPE}) equations, respectively.
Within the macroscopical approach,
the diffusion equation (FPE) is written in terms of
a deformed linear derivative,
while the microscopical approach,
the equations of motion (Langevin), uses
the corresponding dual deformed nonlinear derivative.
This is in complete analogy with the interplay,
reported previously in \cite{Costa-Borges-2014,Costa-Borges-2018},
between the deformed versions of the Schr\"odinger equation and
of the Newton's law obtained in the classical limit.

(iv)
The deformed FPE~(\ref{eq:fpe-inhomogeneous-deformed-derivative}) and the
position-dependent mass Langevin equation~(\ref{eq:masslangevin-pdm-2})
result equivalent to the
nonlinear Langevin equation~(\ref{eq:fullynonlinear-FPE}),
thus guaranteeing the existence of a well-defined stationary solution,
which satisfies the deformed $H$-theorem of
Section \ref{subsec:linear-potential},
and showing a connection between the standard inhomogeneous diffusion
and the one that emerges from a position-dependent mass system.

(v)
The entropy of the system (\ref{eq:deformed-entropy})
is written as the sum of contributions, one from the particles
and one from the medium,
with the latter increasing with deformation,
as illustrated for the case of the confining potential
(Figure \ref{fig:figure4}).
In the context of the van Kampen's diffusion equation
\eqref{eq:van-Kampen-equation}
the entropy contribution of the medium is given in terms of the
position-dependent temperature (equation \eqref{eq:van-Kamppen-entropy}).
For the case of the confining potential and the deformation
\eqref{eq:q-deformation},
the temperature results linear and with the same inverse Gamma distribution
for $f(\beta)$
as in \cite{vanderStraeten}.

(vi)
The solution of the deformed linear FPE for a confining potential
can be obtained from an analogy with the corresponding deformed
linear Schr\"{o}dinger equation (Section \ref{general-solution}).

(vii)
Exponential hyper-diffusion is found for times longer than
the characteristic time, according to the position-dependent mass,
and, consequently, to the deformation parameter, Eq.~(\ref{eq:x-moments}b).

(viii)
Instances addressed in Section \ref{sec:discussion}
point out the potential use of the deformed FPE in different contexts.
Consistency with the van Kampen's inhomogenous diffusion has been established
for the case in which the temperature and the mobility are proportional,
while the position-dependent Langevin equation \eqref{eq:masslangevin-pdm}
in a deformed space
and the superstatistics version of the Langevin equation \eqref{eq:Langevin-superstatistics}
are equivalent.
Two possible realizations of the work fluctuation theorem has been linked
with the diffusion of a position-dependent mass particle,
one of them by averaging the work with the deformation (\ref{eq:q-deformation})
while the other was obtained in terms of the superstatistics approach
in the long term regime.
For the latter we have proposed a modification of the experiment of Wang
by suggesting to employ a temperature and mobility profiles
$T(x)/T_0=\mu(x)/\mu_0=(1+\gamma_qx)$,
in order test power law and exponential decays in the expectation value
of the probability work ratio \eqref{eq:deformed-FDT-confining-potential}
for small and large values of the work $W_\tau$ respectively.
In the general case the van Kampen's equation \eqref{eq:van-Kampen-equation}
can be expressed by \eqref{eq:van-Kampen-general}
in terms of a mixture of deformations
given by the temperature and the mobility.
The van Kampen's FPE along with the superstatistics FPE
and the deformed FPE have the same stationary solution
and satisfy the relationships given by the Table II.

\begin{table}[!htb]\label{tabla2}
 \caption{Structure of the inhomogeneous diffusion
          of the van Kampen's approach, the superstatistics FPE
          and the deformed FPE in the position-dependent mass context.}
\begin{ruledtabular}
\begin{tabular}{ p{0.42\linewidth} | p{0.05\linewidth} | p{0.54\linewidth}}
		deformed FPE (\ref{eq:fpe-inhomogeneous-deformed-derivative})
		with (\ref{eq:choice})
		& \centering $\leftrightarrow$
		& van Kampen FPE (\ref{eq:van-Kampen-equation})
		\\
		\hline
        PDM Langevin equation (\ref{eq:masslangevin-pdm})
        in $y(x)$ (\ref{eq:change-variable})
		&  \vspace{0.5pt} \centering $\leftrightarrow$
		&
		superstatistics
		Langevin Eq.~(\ref{eq:Langevin-superstatistics})
		\newline
		in $y(x)$ (\ref{eq:change-variable})
        \\
        \hline
        deformed FPE (\ref{eq:fpe-inhomogeneous-deformed-derivative})
		& \centering $\nleftrightarrow$
		& superstatistics
		FPE (\ref{superstatistics-FPE})
        \\
        \hline
        superstatistics FPE (\ref{superstatistics-FPE})
		with $T(x)=1/\beta(x)$ and $\mu(x)=\mu_0$
		& \vspace{0.5pt} \centering $\leftrightarrow$
		& \vspace{0.5pt} van Kampen FPE (\ref{eq:van-Kampen-equation})
\end{tabular}
\end{ruledtabular}
\end{table}
Regarding the anomalous diffusion in optical lattices,
the Rayleigh equation for the stationary Wigner function \cite{Lutz}
can be expressed as a deformed FPE in a deformed momentum space $p_{[1/D]}$,
with $D(p)$ the diffusion coefficient.

There is an equivalence between the deformed space of the position-dependent
mass system, the heterogeneity of the environment and the superstatistics,
which could potentially be used to study problems in these areas.

Table I uses the two deformed derivatives (one linear and one nonlinear)
for which the $q$-exponential is the eigenfunction.
In order to complete the scheme,
it is still missing the development of deformed versions
of FPE and Sch\"odinger equation using their dual derivatives,
i.e., those whose the deformed derivative of the $q$-logarithm of $u$ is $1/u$:
$\widetilde{\mathcal{D}}_{q} f(u)$
(Eq.~(\ref{eq:borges-nonlinear-derivative-q-diff}))
and
$\mathfrak{D}_{q} f(u)$,
(Eq.~(\ref{eq:nobre-linear-derivative})).

The linear deformation of the FPE addressed in this paper
does not formally departs from Boltzmann-Gibbs statistical mechanics,
within the deformed space
(see Eqs.~(\ref{eq:total-entropy}), (\ref{eq:deformed-entropy})).
It is interesting to explore the consequences of nonlinear deformations
to identify which case leads to a nonextensive statistical mechanics scenario
described by $S_q$ entropy.
Besides,
other deformed algebras could be employed,
for instance within the context of relativistic statistical mechanics
\cite{Kaniadakis-2002}
as well as those from entropic information generalizations \cite{Beck-entropy,Rodriguez-2019},
thus leading to different deformations of the FPE.

\begin{acknowledgments}
I.\ S.\ G.\ and E.\ P.\ B.\ acknowledge support from National Institute of
Science and Technology for Complex Systems (INCT-SC).
I.\ S.\ G.\ also acknowledges support from
Coordena\c{c}\~ao de Aperfei\c{c}oamento de Pessoal de N\'ivel Superior (CAPES)
and Conselho Nacional de Desenvolvimento Cient\'ifico e Tecnol\'ogico
(CNPq -- Postdoctoral Fellowship 159799/2018-0),
Brazilian agencies.
\end{acknowledgments}



\end{document}